\documentclass[jcp,aip,floatfix,preprintnumbers,superscriptaddress,preprint]{revtex4-1}
\usepackage{graphicx}
\usepackage{epsfig}
\usepackage{epstopdf}
\usepackage{amsmath}
\usepackage{amssymb}
\usepackage[subrefformat=parens,labelformat=parens]{subfig}
\usepackage{float}
\usepackage{amssymb}
\usepackage{color}
\usepackage{mathtools}

\graphicspath{{./fig/}}

\newcommand{\beginsupplement}{%
        \setcounter{table}{0}
        \renewcommand{\thetable}{S\arabic{table}}%
        \setcounter{figure}{0}
        \renewcommand{\thefigure}{S\arabic{figure}}%
     }

\newcommand{\beq}{\begin{equation}}
\newcommand{\eeq}{\end{equation}}
\newcommand{\bea}{\begin{eqnarray}}
\newcommand{\eea}{\end{eqnarray}}

\newcommand{\pospart}[1]{\text{max}\left(#1,0\right)}

\newcommand{\nene}[1]{#1\pm 1}

\newcommand{\mean}[1]{\overline{#1}}
\newcommand{\changed}[1]{#1}

\begin{document}


\title{Optimisation of Simulations of Stochastic Processes by Removal of Opposing Reactions}


\author{Fabian Spill}
\affiliation{Department of Biomedical Engineering, Boston University, 44 Cummington Street, Boston MA 02215, USA}
\affiliation{Department of Mechanical Engineering, Massachusetts Institute of Technology,\\ 77 Massachusetts Avenue, Cambridge, MA 02139, USA}
\author{Philip K. Maini}
\author{Helen Byrne}
\affiliation{Wolfson Centre for Mathematical Biology, Mathematical Institute, University of Oxford, Oxford OX2 6GG, UK}

\date{\today}

\begin{abstract}
Models invoking the chemical master equation are used in many areas of science, and, hence, their simulation is of interest to many researchers. The complexity of the problems at hand often requires considerable computational power, so a large number of algorithms have been developed to speed up simulations. However, a drawback of many of these algorithms is that their implementation is more complicated than, for instance, the Gillespie algorithm, which is widely used to simulate the chemical master equation, and can be implemented with a few lines of code. Here, we present an algorithm which does not modify the way in which the master equation is solved, but instead modifies the transition rates, and can thus be implemented with a few lines of code. It works for all models in which reversible reactions occur by replacing such reversible reactions with effective net reactions. Examples of such systems include reaction-diffusion systems, in which diffusion is modelled by a random walk. The random movement of particles between neighbouring sites is then replaced with a net random flux. Furthermore, as we modify the transition rates of the model, rather than its implementation on a computer, our method can be combined with existing algorithms that were designed to speed up simulations of the stochastic master equation. By focusing on some specific models, we show how our algorithm can significantly speed up model simulations while maintaining essential features of the original model.
\end{abstract}


\maketitle

\section{Introduction}
Stochastic models are used in many areas of physics, chemistry and biology to describe random fluctuations associated with uncertainties in the external environment and the intrinsic discreteness of the species of interest, which could be individual atoms, molecules, cells or animals ~\cite{van1992stochastic,gardiner2010stochastic}. The chemical master equation is commonly used to determine the time evolution of the probability density function which describes the current state of the system of interest. Such a state could, for instance, be defined by the current number of molecules of each kind in a model of chemical reactions ~\cite{gillespie2013perspective,de2015effects,dobramysl2015particle}, the number of animals of each species in an ecological model ~\cite{black2012stochastic}, or the number of cells in an animal tissue ~\cite{spill2014mesoscopic,sturrock2014role,guerrero2015invasion}. The problem which remains is then to solve for the time-evolution of the probability density function.

As analytic solutions to the master equations are rarely obtainable, one typically aims to simplify the master equation, (e.g. by the van Kampen expansion ~\cite{van1992stochastic}), or one numerically computes solutions of the master equation. A commonly used algorithm for such simulations is the Gillespie algorithm ~\cite{gillespie1976general,gillespie1977exact}. It is exact in the sense that every reaction is taken into account, and no stochastic information is lost in the simulation process. However, this makes it computationally expensive, \changed{as for each event several potentially expensive computational steps, such as the simulation of random numbers, the recalculation of propensities or the identification of the next occurring reaction need to be performed}. Additionally, in many cases, the large number of elementary reactions involved makes the use of the conventional Gillespie algorithm practically unfeasible.

For this purpose, a number of algorithms have been developed in order to speed up the simulation of the master equation. Some of these are hybrid methods that switch from a stochastic to the corresponding mean field model ~\cite{gibson2000,moro2004,hellander2007hybrid,henzinger2010hybrid,spill2015hybrid} in regions of space or time where the latter are reasonable approximations to the stochastic equations. Others, such as the tau-leap method ~\cite{gillespie2001approximate,gillespie2003improved,gillespie2007stochastic,
cao2006efficient,fu2013time,wu2015adaptive}, simulate several reactions within one time-step. Related work focuses on more efficient simulation of stochastic models not based on the chemical master equation ~\cite{flekkoy2001coupling,alexander2002algorithm,alexander2005algorithm,li2012spatially,flegg2012,franz2013,
flegg2015convergence,robinson2014adaptive,sanft2015constant,yates2015pseudo}.

A common feature of the aforementioned methods is that their implementation is typically more complex than for the Gillespie algorithm, which can be implemented with a few lines of code. In this paper, we introduce an algorithm which does not affect the solution method of the master equation, but rather changes the transition rates of the master equation itself. As a result, existing algorithms (and their optimisations) can be used, and only the definition of the transition rates needs to be changed. The idea is to group together reversible reactions. For example, a particle which is created and subsequently annihilated before it reacts with any other particle has no observable effect. Likewise, in a stochastic diffusion problem on a lattice, if two particles of the same type swap sites, this has no observable effect on the state of the system, provided no reactions happen while the particles switch locations. Our algorithm involves combining transition rates for reactions which have opposing effects to produce effective rates that describe the net effect of the two reactions. The model, with the modified transition rates, is designed such that the mean behaviour is the same as that of the original stochastic model, but typically stochastic effects such as variance are reduced. Hence, we suggest that the modified transition rates are used when the number of particles involved in the reaction exceeds a threshold value chosen such that stochastic effects corresponding to the reaction of the transition rate which we modify can be ignored. This is similar to methods such as ~\cite{spill2015hybrid} which replace parts of a stochastic model with the mean-field description of the same model. In the mean-field limit of a stochastic model, the dynamics of two opposing reactions is also only taken into account in a net, effective way. However, unlike the mean-field limit, our method is based entirely on the chemical master equation. Hence, it is easier to implement than most other hybrid methods, as the modifications to our algorithm are baseed on the transition rates and, hence, can be easily combined with transition rates of other reactions which are not optimised. \changed{Furthermore, in contrast to other methods~\cite{franz2013,spill2015hybrid}, which approximate a stochastic process in some regions by its mean-field description, our method does not encounter technical challenges at the interface of the stochastic and mean field domains.}

\changed{
The remainder of the paper is organised as follows. We discuss a simple birth-death process in section \ref{sec:singleReaction}. This example serves to introduce our algorithm, and can have applications in complex reaction networks where at least some reactions are reversible. In section \ref{sec:stochasticDiffusion} we focus on the stochastic diffusion of particles randomly migrating on a lattice. This is a good test of our algorithm since when two particles move in opposite directions, the system returns to its original state. We show that our model of stochastic diffusion accurately preserves stochastic properties such as first-passage times. Our study of stochastic diffusion is also important for understanding how our algorithm will perform for the more general case of reaction-diffusion systems. For instance, when diffusion occurs on a faster timescale relative to reactions, many diffusive events may occur which do not affect the reactions, but slow down simulations of the system. In section \ref{sec:Fisher} we focus on the stochastic Fisher-Kolmogorov system to how that our algorithm can increase simulation speed significantly while maintaining essential stochastic features, such as the modification of the wave speed due to stochastic effects. In section \ref{sec:Min}, we study a one-dimensional model of Min oscillations. This multi-species example illustrates how our algorithm performs when some molecular species are present at low copy numbers, leading to stochastic effects, while others are present in high numbers, making simulations slow. Again, our method can markedly improve simulation speed, while preserving the statistical distributions of species with low copy numbers.}
%
\section{Creation and Annihilation Process}\label{sec:singleReaction}
We consider a spatially averaged model in which particles of type A can either divide, or annihilate when they hit another particle of the same type:
\begin{align}
\begin{split}
A &\xrightarrow[]{\lambda} 2A,\\
2A &\xrightarrow[]{\mu} A
\end{split}
\end{align}
%
Here, $\lambda$, and $\mu$ are rate constants associated with division and annihilation, i.e. for small times $t$, $\lambda t + \mathcal{O}(t^2)$ is the probability that a randomly chosen particle divides, and $2\mu t + \mathcal{O}(t^2)$ is the probability that one of two randomly chosen particles annihilates the other. Such a model is the simplest to which our algorithm can be applied, as it consists of only two reactions which have equal and opposite effects. Our algorithm replaces the creation and annihilation reactions with a net reaction which, depending on whether creation or annihilation is more likely, will itself be a creation or annihilation reaction with a rate given by the difference of the rates of the original model.

\begin{figure}[h!]
\includegraphics[width=0.66\linewidth]{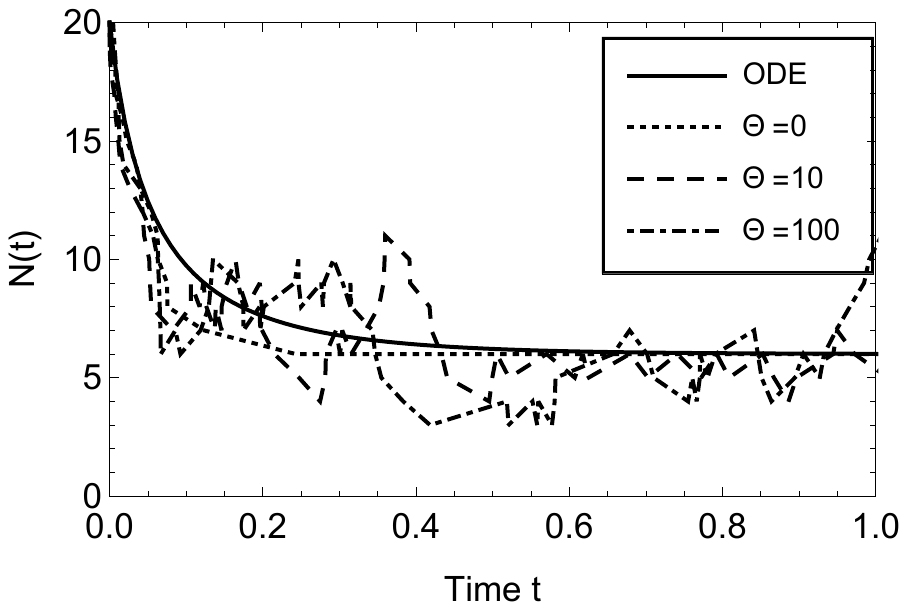}
\caption{\changed{Birth-death process with parameters $\lambda=5$, $\mu =1$ and $N=20$ at $t=0$, simulated with the original stochastic model \eqref{eq:transitionRatesBirthDeath} (obtained by setting the threshold $\Theta=100$), the difference model \eqref{eq:transitionRatesBirthDeathOptimised} ($\Theta=0$) and the conditional difference model \eqref{eq:transitionRatesBirthDeathMixed} with $\Theta=10$, and the mean-field ODE \eqref{eq:meanFieldBirthDeath}.}
\label{fig:BirthDeath}} 
\end{figure}

We model this process with the chemical master equation, denoting by $P(N,t)$ the probability density function  for $N$ particles to be present in the system at time $t$:
\begin{align}
\label{eq:MasterEquationBirthDeath}
 \frac{dP(N,t)}{dt} &= \left((E^--1)\mathcal{T}_{N+1|N} + (E^{+1}-1)\mathcal{T}_{N-1|N}\right)P(N,t),\quad N \in \mathbb{N}
 \end{align}
Here, $E^\pm$ are shift operators that shift the number of particles $N$ by $\pm 1$, and $\mathcal{T}_{N+1|N}$ ($\mathcal{T}_{N-1|N}$) denote the transition rates for the creation (annihilation) of a particle, \changed{so that}
\begin{align}
\begin{split}
\label{eq:transitionRatesBirthDeath}
\mathcal{T}_{N+1|N} &= \lambda N,\\
\mathcal{T}_{N-1|N} &= \mu N (N-1).
\end{split}
\end{align}
We note that $\mathcal{T}_{0|1} = \mathcal{T}_{-1|0} = 0$, so the model ensures that neither extinction nor negative particle numbers can occur, if $N$ is initially positive. \changed{From equations \eqref{eq:MasterEquationBirthDeath}, \eqref{eq:transitionRatesBirthDeath} we can see that if $\lambda N > \mu N (N-1)$, then a new particle is more likely to appear than to disappear. Conversely, if $\lambda N < \mu N (N-1)$, particles will be more likely to disappear. We thus define a second stochastic process with the same master equation \eqref{eq:MasterEquationBirthDeath}, but transition rates given by}
\begin{align}\begin{split}\label{eq:transitionRatesBirthDeathOptimised}
\mathcal{T}_{N+1|N} &= \pospart{\lambda N - \mu N (N-1)},\\
\mathcal{T}_{N-1|N} &= \pospart{\mu N (N-1) - \lambda N}.
\end{split}\end{align}
\changed{It is straighforward to show that, if $\mean{N^2}\approx \mean{N}^2$, which can be motivated, for instance, by the van-Kampen expansion ~\cite{van1992stochastic} in the limit of large particle numbers, then the model with transition rates \eqref{eq:transitionRatesBirthDeathOptimised} leads to the mean-field equation}
\begin{align}\label{eq:meanFieldBirthDeath}
\frac{d\mean{N}}{dt} \approx \lambda\mean{N}-\mu\mean{N}(\mean{N}-1),\quad \mean{N} =\sum_N NP(N,t),
\end{align}
\changed{
which is identical to that for the model with transition rates \eqref{eq:transitionRatesBirthDeath}.} However, the behaviours of \eqref{eq:transitionRatesBirthDeath} and \eqref{eq:transitionRatesBirthDeathOptimised} for low particle numbers and low ratios of $\frac{\lambda}{\mu}$ are quite different (see also Figure \ref{fig:BirthDeath} and the discussion below). We note that for a given state, transition rates \eqref{eq:transitionRatesBirthDeathOptimised} are always always bounded above by those in \eqref{eq:transitionRatesBirthDeath}. Therefore, a global change in state will typically involve fewer stochastic events in \eqref{eq:transitionRatesBirthDeathOptimised} than in \eqref{eq:transitionRatesBirthDeath} and be faster to simulate with the Gillespie algorithm. We henceforth refer to \eqref{eq:transitionRatesBirthDeathOptimised} as the difference model, as it is constructed by taking the differences of reversible reactions in the original, exact model \eqref{eq:transitionRatesBirthDeath}. \changed{Having established that for large particle numbers \eqref{eq:transitionRatesBirthDeathOptimised} and \eqref{eq:transitionRatesBirthDeath} behave similarly, and that for small particle numbers they do not, we now introduce a model which conditionally switches between transition rates \eqref{eq:transitionRatesBirthDeath} and \eqref{eq:transitionRatesBirthDeathOptimised}:}
\begin{align}\begin{split}\label{eq:transitionRatesBirthDeathMixed}
\mathcal{T}_{N+1|N} &= \begin{cases}
\pospart{\lambda N - \mu N (N-1)}\quad &\text{if } N\geq\Theta,\\
\lambda N \quad&\text{otherwise},
\end{cases},
\\
\mathcal{T}_{N-1|N} &= 
\begin{cases}
\pospart{\mu N (N-1)-\lambda N }\quad &\text{if } N\geq\Theta\\
\mu N (N-1)\quad&\text{otherwise.}
\end{cases}
\end{split}\end{align}
Here, we have introduced a threshold $\Theta$ \changed{such that when  $N < \Theta$ we use the exact model \eqref{eq:transitionRatesBirthDeath}, and when $N\geq\Theta$ we approximate  \eqref{eq:transitionRatesBirthDeath} by the difference model \eqref{eq:transitionRatesBirthDeathOptimised}. We will refer to \eqref{eq:transitionRatesBirthDeathMixed} as the conditional difference model. If $\Theta\to\infty$ we recover \eqref{eq:transitionRatesBirthDeath}, whereas \eqref{eq:transitionRatesBirthDeathOptimised} is obtained for $\Theta=0$. Note that while \eqref{eq:transitionRatesBirthDeathMixed} leads to the same mean-field equation as \eqref{eq:transitionRatesBirthDeath}, its stochastic properties are, generally, different. We can view \eqref{eq:transitionRatesBirthDeathMixed} as an approximation to \eqref{eq:transitionRatesBirthDeath}, and the challenge is to choose $\Theta$ so that simulation results are in close agreement with \eqref{eq:transitionRatesBirthDeath}.}
Figure \ref{fig:BirthDeath} shows a typical simulation of the birth-death process with parameters $\lambda=5$, $\mu =1$ and $N=20$ at $t=0$. We compare results generated from the \changed{exact model \eqref{eq:transitionRatesBirthDeath}, the difference model \eqref{eq:transitionRatesBirthDeathOptimised}, the conditional difference model \eqref{eq:transitionRatesBirthDeathMixed} with a threshold value $\Theta=10$ and the corresponding ODE model, equation \eqref{eq:meanFieldBirthDeath}. We see that the behaviour of the conditional difference model is qualitatively similar to that of the original model. \changed{By contrast, the number of particles in the difference model decays randomly until it reaches $N=\frac{\lambda}{\mu} + 1 = 6$, which is an absorbing state in the difference model.} Hence, the difference model behaves qualitatively differently from the original, exact model and the conditional difference model when $N\approx 6$. It behaves more like the mean-field ODE solution because of the absorbing state, which is not present in the original model \eqref{eq:transitionRatesBirthDeath}, and when $N>6$, particles in the difference model can only decay, i.e. cannot be created. In supplementary Figure \ref{fig:BirthDeathDistribution}, we show the distribution of numbers for two thresholds, $\Theta = 0,10$, at two times, shortly after the start of the simulations, $t=0.01$, and at the long time, $t=10$, when the solution is close to its steady state. In each case we compare the results to those of the exact model, which is obtained by fixing $\Theta=100$. We note that the distributions at $t=0.01$ are similar for all cases, but when the threshold is less than the initial condition, the distribution is bounded above by the initial condition. At $t=10$, we notice that for $\Theta=0$, the distribution approximates the delta distribution, whereas when $\Theta=10$, it closely matches that for the exact model. We explain those results by noting that in the exact model, particle numbers exceed $N=10$ only with low probability.

A more detailed comparison between the exact model and its approximation by the difference and conditional difference models will be performed in the sections that follow, where we consider the examples of diffusion and reaction-diffusion problems.}
\section{Stochastic Diffusion Models}\label{sec:stochasticDiffusion}


%
We now consider stochastic versions of the one-dimensional diffusion equation,
\begin{align}\begin{split}\label{eq:Diffusion}
\frac{\partial n}{\partial t} = D\frac{\partial^2 n}{\partial x^2},
\end{split}\end{align}
defined for $x\in[0,L]$ and $t\geq0$, and which is supplemented with boundary conditions which we do not specify here for generality. In equation \eqref{eq:Diffusion}, $D>0$ is the constant diffusion coefficient, and the concentration $n=n(x,t)$ denotes the number of particles per unit length. We keep all quantities in dimensional form to better understand the physics. We discretise the spatial domain into $k_{max}$ equally spaced compartments of size $h$. We denote by $N_k(t)$ the number of particles in compartment $k$ at time $t$, dropping the time dependence when no confusion is possible, and consider the master equation 
\begin{align}\begin{split}
\label{eq:generalMasterEquation}
 \frac{dP(N_{j},t)}{dt} &= \sum_k\sum_{l=\nene{k}}(E^+_kE^-_l-1)\mathcal{T}_{N_k-1,N_l+1|N_k,N_l}P(N_{j},t),
 \end{split}\end{align}
where $P(N_{j},t)$ denotes the probability density function that there are $N_{j}$ particles in compartment $j$ at time $t$, for $j=1,\dots,k_{max}$. Furthermore, $\mathcal{T}_{N_k-1,N_l+1|N_k,N_l}$ denotes the transition rate from a state in which there are $N_k,N_l$ particles in compartments $k$ and $l$, respectively, to one in which there are $N_k-1$ and $N_l+1$ particles in those compartments, while particle numbers in all other compartments remain constant. Finally, the operators $E^\pm_k$ are defined by increasing (decreasing) $N_k$ by one. A simple random walk model is defined by the transition rates 
\begin{align}\begin{split}\label{eq:transitionRatesRandomWalk}
\mathcal{T}_{N_k-1,N_{k\pm 1}+1|N_k,N_{k\pm 1}} &= \frac{D}{h^2}N_k.
\end{split}\end{align}
Here, $D$ is the diffusion coefficient that appears in equation \eqref{eq:Diffusion}. \changed{Henceforth, we refer to the model with transition rates \eqref{eq:transitionRatesRandomWalk} as the exact model of stochastic diffusion}. As the transition rate is proportional to the number of particles in the outgoing box $k$, this indicates that the random walkers do not interact. The mean field equations associated with this stochastic model are given by 
\begin{align}\begin{split}\label{eq:meanEquation}
\frac{\partial \mean{N_k}}{\partial t} = \frac{D}{h^2}
\left(\mean{N_{k+1}}-2\mean{N_k}+\mean{N_{k-1}}\right).
\end{split}\end{align}
Equation \eqref{eq:meanEquation} can be viewed as an explicit finite-difference approximation of the diffusion equation \eqref{eq:Diffusion} if we identify the densities with the particle numbers via $N_k(t) = h n(x,t), x= k h$. We now propose the following, alternative transition rates for a model of stochastic diffusion: 
\begin{align}\begin{split}\label{eq:transitionRatesDifference}
\mathcal{T}_{N_k-1,N_{k\pm 1}+1|N_k,N_{k\pm 1}} &= \frac{D}{h^2}\pospart{N_k-N_{k\pm 1}}.
\end{split}\end{align}
Here, a random walker moves from compartment $k$ to compartment $k\pm1$ only if the number of particles in the outgoing box $k$ exceeds the number in the incoming box $k\pm1$. Transition rates \eqref{eq:transitionRatesDifference} give rise to the same mean field equations \eqref{eq:meanEquation} as transition rates \eqref{eq:transitionRatesRandomWalk} and, hence, also lead to the diffusion equation in the continuum limit. Indeed, the net average flux between neighboring compartments implied by transition rates \eqref{eq:transitionRatesDifference} is the same as that implied by \eqref{eq:transitionRatesRandomWalk}. \changed{However, we can expect that the noise associated with \eqref{eq:transitionRatesRandomWalk} will be larger, since the transition rates \eqref{eq:transitionRatesDifference} are bounded above by the transition rates \eqref{eq:transitionRatesRandomWalk}. In situations where noise changes the system dynamics, \eqref{eq:transitionRatesDifference} might not be suitable. For this reason, we consider a combination of 	\eqref{eq:transitionRatesRandomWalk} and \eqref{eq:transitionRatesDifference}. We introduce a non-negative threshold $\Theta$  and the conditional difference model}
\begin{align}\begin{split}\label{eq:transitionRatesMix}
\mathcal{T}_{N_k-1,N_{k\pm 1}+1|N_k,N_{k\pm 1}} &= 
\begin{cases}
\frac{D}{h^2}\pospart{N_k-N_{k\pm 1}}\quad \text{if } N_k,N_{k\pm1}\geq\Theta,\\
\frac{D}{h^2}N_k\quad\text{otherwise}.
\end{cases}
\end{split}\end{align}
We motivate \eqref{eq:transitionRatesMix} by noting that \eqref{eq:transitionRatesRandomWalk} \changed{represents the exact model of the random walkers, and should be used when particle numbers are low, since typically then relative noise levels are significant. When particle numbers are high, i.e. $N_k, N_{k\pm1}\geq \Theta$, we approximate \eqref{eq:transitionRatesRandomWalk} by \eqref{eq:transitionRatesDifference}, as for high particle numbers relative noise levels are typically low. We note that the mean of \eqref{eq:transitionRatesDifference} and \eqref{eq:transitionRatesRandomWalk} are both given by \eqref{eq:meanEquation}, and $\Theta$ should be scaled with the lattice constant $h$. In particular, when the lattice constant is small, even if the total number of particles present in the system is large, the number of particles per compartment may not be large. In such cases the threshold would need to be reduced, but with caution since the application of \eqref{eq:transitionRatesDifference} might lead to restrictive suppression of fluctuations.

We now aim to identify situations for which transition rates \eqref{eq:transitionRatesRandomWalk}, \eqref{eq:transitionRatesDifference} and \eqref{eq:transitionRatesMix} yield similar behaviour, and when they do not.
}

\subsection{First-Passage Time problems}\label{sec:firstPassageTime}
\changed{In many transport problems, it is important to know when a particle has first reached a certain site.} We thus begin by investigating a first-passage time problem. We also refer the reader to ~\cite{bezzola2014exact} for more studies of first-passage time problems. We suppose that initially $N^0$ particles are located at the left boundary compartment $k=1$, and all other compartments are empty. We impose Dirichlet boundary conditions, so that $N_1=N^0$ and $N_{k_{max}}=0$ at all times, and the lattice constant is $h=1$. We define the first-passage time at site $k$ to be the time at which a particle first reaches compartment $k$. 
\begin{figure}[h!]
\includegraphics[width=0.68\linewidth]{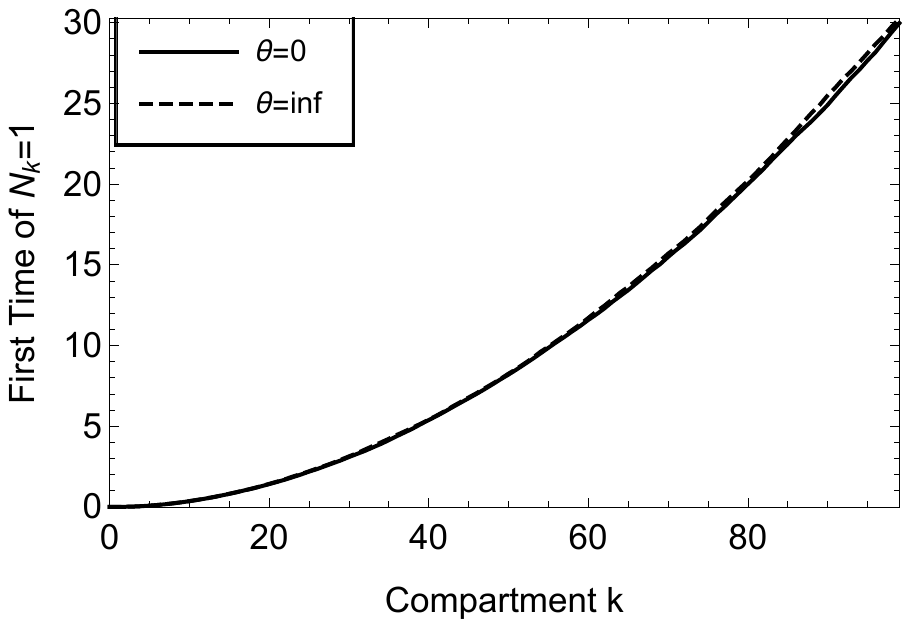}
\caption{First-passage times for $N^0=1000$ random walkers starting at compartment $k=1$ at time $0$, with $D=10$. We present the average first-passage time for conventional random walkers based on the exact model defined by \eqref{eq:transitionRatesRandomWalk} (solid line) and for the random walkers based on the difference model \eqref{eq:transitionRatesDifference} (dashed line). The results are averaged of $1024$ simulations and are in excellent agreement.
\label{fig:FirstPassageTimes}} 
\end{figure}
In Figure \ref{fig:FirstPassageTimes} we compare the average first-passage times obtained by averaging $1024$ realisations of the \changed{exact random walk model \eqref{eq:transitionRatesRandomWalk}, which coincides with the conditional difference model when $\Theta\to \infty$, and the random walk based on the difference model \eqref{eq:transitionRatesDifference}, which coincides with the conditional difference model when $\Theta = 0$}. As expected, there is excellent agreement between the two models, because the first-passage times represent the time at which a single particle enters an empty compartment, and the transition rates for entry into an empty compartment are identical for \eqref{eq:transitionRatesRandomWalk} and \eqref{eq:transitionRatesDifference}.

\begin{figure}[h!]
\includegraphics[width=0.68\linewidth]{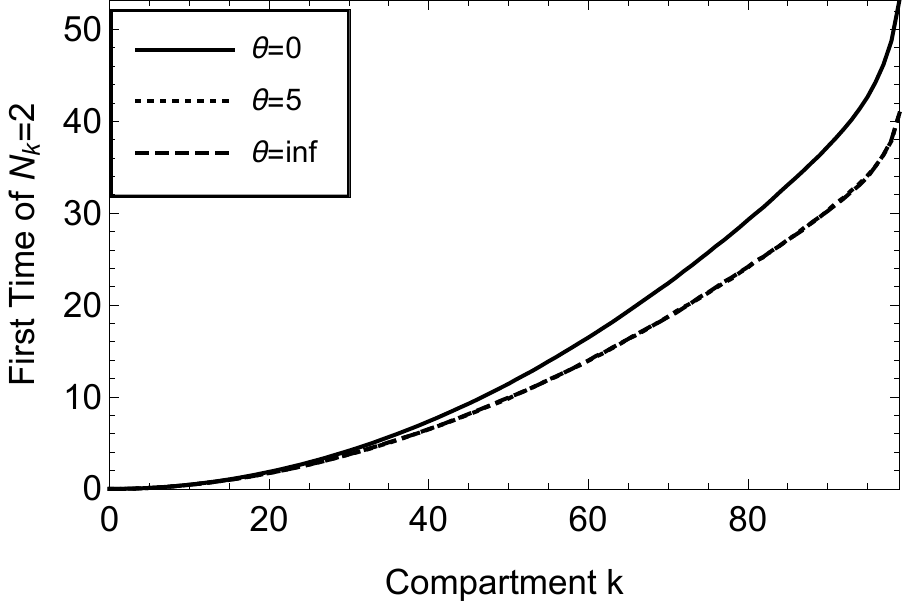}
\caption{ The first time when a compartment $k$ contains two particles is shown as a function of $k$ for \changed{the exact random walk \eqref{eq:transitionRatesRandomWalk}(solid line, $\Theta=\infty$), the random walk based on the difference model \eqref{eq:transitionRatesDifference}(dotted line, $\Theta = 0$) as well as the conditional difference model \eqref{eq:transitionRatesMix} (dashed line, $\Theta = 5$)}. Each plot is obtained by averaging $1024$ simulations, with $D=10$ and $N^0=1000$ random walkers at time $0$ in compartment $k=1$. The graphs for the exact random walk and the conditional difference model with $\Theta=5$ coincide.
\label{fig:FirstPassageTwoHits}} 
\end{figure}
When studying problems such as a reaction-diffusion system in which two, or more, particles of the same species need to be present at the same time and place, we should also establish whether changes in the transition rates of the random walk affect the time until several particles reach the same location. In such situations, we are concerned with the distribution of first times at which two particles are present in the same compartment. The first-passage times for two particles are shown in Figure \ref{fig:FirstPassageTwoHits}. We notice that the difference model \eqref{eq:transitionRatesDifference} predicts first passage times which are slightly longer than those predicted by the conditional difference model \eqref{eq:transitionRatesMix} with a threshold $\Theta=5$, whose first-passage times are indistinguishable from those of the exact random walk model \eqref{eq:transitionRatesRandomWalk}. This is because when one particle is present in a compartment, the rate at which a second particle enters that compartment is lower for the difference model, but identical for the conditional difference model when $\Theta\geq 2$ and the exact random walk model. Hence, when the system of interest involves reactions with two or more particles, the threshold needs to be set suitably high. \changed{In supplementary Figure \ref{fig:DiffusionDistribution}, we present histograms of the number of particles in a given box at a given time, for the same scenario as above, comparing the distributions associated with different thresholds to those associated with the exact model $\Theta=\infty$. We find that, in general, lower thresholds lead to narrower distributions centered around the same mean, and that as the threshold levels are increases, the distribution approaches that obtained from the exact model. In supplementary Figure \ref{fig:DiffusionDistributionMoments} we confirm that, as the threshold increases, the standard deviations of the exact model is recovered in a switch-like manner.}

\begin{table}[h]
\begin{tabular}{|c|c|c|}
\multicolumn{1}{c}{}& \multicolumn{1}{c}{Number of Gillespie Events} \\ \hline
 &  $N^0=100$ & $N^0=1000$ \\ \hline
 \changed{Exact random walk} \eqref{eq:transitionRatesRandomWalk}& $1.21 \times 10^6$ & $7.97\times 10^6$ \\
Difference model \eqref{eq:transitionRatesDifference}&	$4.68\times 10^4$ & $3.04\times 10^5$ \\
\changed{Conditional difference model} \eqref{eq:transitionRatesMix}, $\Theta=5$&	$7.35\times 10^4$ & $3.23\times 10^5$ \\ 
\changed{Conditional difference model}\eqref{eq:transitionRatesMix}, $\Theta=10$& $1.06 \times 10^5$	& $4.88\times 10^5$ \\ \hline
 \end{tabular}
  \caption{Comparison of the average number of Gillespie events needed to simulate the first-passage time problem, shown in Figure \ref{fig:FirstPassageTimes}, for the exact random walk model \eqref{eq:transitionRatesRandomWalk}, the difference model \eqref{eq:transitionRatesDifference} and the \changed{conditional difference model} \eqref{eq:transitionRatesMix} with two different thresholds $\Theta = 5,10$. The simulations were stopped when every lattice site had been visited at least once by at least one particle.}
 \label{tab:NumberGillespieEvents}
 \end{table}

In Table \ref{tab:NumberGillespieEvents} we show the average number of stochastic events needed to ensure that all lattice sites have been visited at least once by at least one particle. We fix $D=1$ and average over $1024$ simulations. We see that the random walk difference model \eqref{eq:transitionRatesDifference} needs only $4\%$ of the number of events that the conventional random walk model \eqref{eq:transitionRatesRandomWalk} needs, with the conditional difference model \eqref{eq:transitionRatesMix} offering similar performance gains, although these improve as the threshold $\Theta$ is reduced. We remark also that the time to simulate a stochastic model with the conventional Gillespie algorithm is proportional to the total number of events that occur, since we repeat the same steps when simulating a single event. We conclude that the \changed{conditional difference model} preserves the essential stochastic features of the \changed{exact random walk model} while being considerably faster to simulate.
%
%

\section{Reaction-Diffusion Models}
\subsection{Stochastic Fisher-Kolmogorov Equation}\label{sec:Fisher}

\changed{We now investigate a stochastic version of the Fisher-Kolmogorov equation. This equation can be viewed as a spatial-resolved birth-death process, in which identical particles move in a diffusive manner, divide and/or annihilate. The stochastic model is defined by the master equation}
\begin{align}\begin{split}
\label{eq:MasterEquationFisher}
 \frac{dP(N_{j},t)}{dt} &= \sum_k\sum_{l=\nene{k}}(E^+_kE^-_l-1)\mathcal{T}_{N_k-1,N_l+1|N_k,N_l}P(N_{j},t) \\ 
  &+\sum_k\left((E^-_k-1)\mathcal{T}_{N_k+1|N_k} + (E^{+1}_k-1)\mathcal{T}_{N_k-1|N_k}\right)P(N_{j},t),
 \end{split}\end{align}
where the birth and death transition rates are defined by
\begin{align}\begin{split}\label{eq:transitionRatesFisher}
\mathcal{T}_{N_k+1|N_k} &= \lambda N_k,\\
\mathcal{T}_{N_k-1|N_k} &= \frac{\lambda}{ \Omega h} N_k (N_k-1),
\end{split}\end{align}
and the random walk transition rates are defined by either \eqref{eq:transitionRatesRandomWalk}, \eqref{eq:transitionRatesDifference} or \eqref{eq:transitionRatesMix}. \changed{In \eqref{eq:transitionRatesFisher}, $\lambda$ is a growth rate and $h\Omega$ is a measure of the carrying capacity of a single compartment, and, as before, $h$ denotes the lattice constant. Consequently, if $N_k$ is larger than $h\Omega$, then the annihilation process dominates the creation process. If we assume, as for the case of the homogeneous birth-death process, that the particle number is sufficiently large that a van-Kampen limit applies, then this stochastic model results in mean field equations of the form}
\begin{align}\begin{split}\label{eq:meanEquationFisher}
\frac{\partial \mean{N_k}}{\partial t} = \frac{D}{h^2}\left(\mean{N_{k+1}}-2\mean{N_k}+\mean{N_{k-1}}\right) + \lambda\mean{N_k}\left(1-\frac{\mean{N_k}}{\Omega h}\right).
\end{split}\end{align}
Then, setting \changed{$n(x,t)h = N_k(t)$}, $x = kh$, as before, we obtain the Fisher-Kolmogorov equation in the continuum limit:
\begin{align}\begin{split}\label{eq:Fisher}
\frac{\partial n}{\partial t} = D\frac{\partial^2 n}{\partial x^2} + \lambda n\left(1-\frac{n}{\Omega}\right)
\end{split}\end{align}
\changed{An interesting feature of equation \eqref{eq:Fisher} is that it generates stable travelling wave solutions which propagate with a constant speed of $2\sqrt{D\lambda}$ (see supplementary Figure \ref{fig:FisherDeterministic} for an example of a travelling wave obtained from equation \eqref{eq:Fisher}). The wave speed of the stochastic Fisher-Kolmogorov equation differs from that of the corresponding PDE ~\cite{breuer1994fluctuation}. Hence, the challenge for our algorithm is to reproduce this shift in wave speed, while being faster to simulate than the exact model. To this end, we now investigate travelling wave propagation in the stochastic Fisher-Kolmogorov equation, when the random walk part is modelled by the exact model \eqref{eq:transitionRatesRandomWalk} and by the difference model \eqref{eq:transitionRatesDifference}. For clarity, we do not apply our difference method to the reaction rates \eqref{eq:transitionRatesFisher}, since otherwise it will be less clear if the observed differences from using the difference model are due to the reaction of the diffusion parts of the model. Initial conditions are chosen so that the travelling wave front resembles that from the PDE. We impose Dirichlet boundary conditions so that $N_1=\Omega$ and $N_{k_{max}}=0$ for all times. We focus our analysis on the propagation of fully formed travelling wave solutions whose leading edge is at a distance from the boundary, so that the initial and boundary conditions do not influence the propagating wave.}
\begin{figure}[h!]
\subfloat[Conventional Random Walk]
{\label{fig:FisherTravellingWavesNorm}
\includegraphics[width=0.48\linewidth]{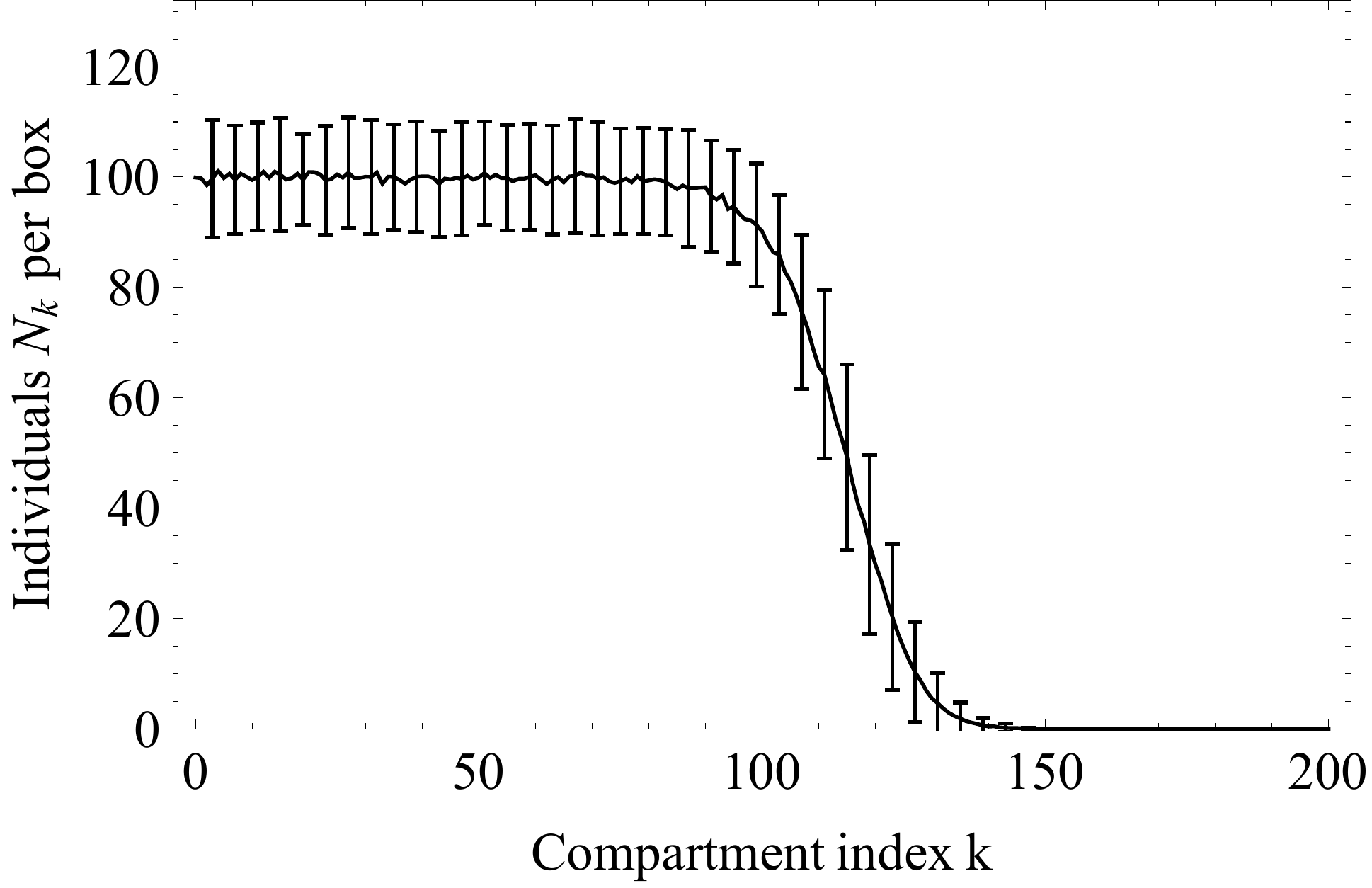}
}
\subfloat[Difference Random Walk]
{\label{fig:FisherTravellingWavesOpt}
\includegraphics[width=0.48\linewidth]{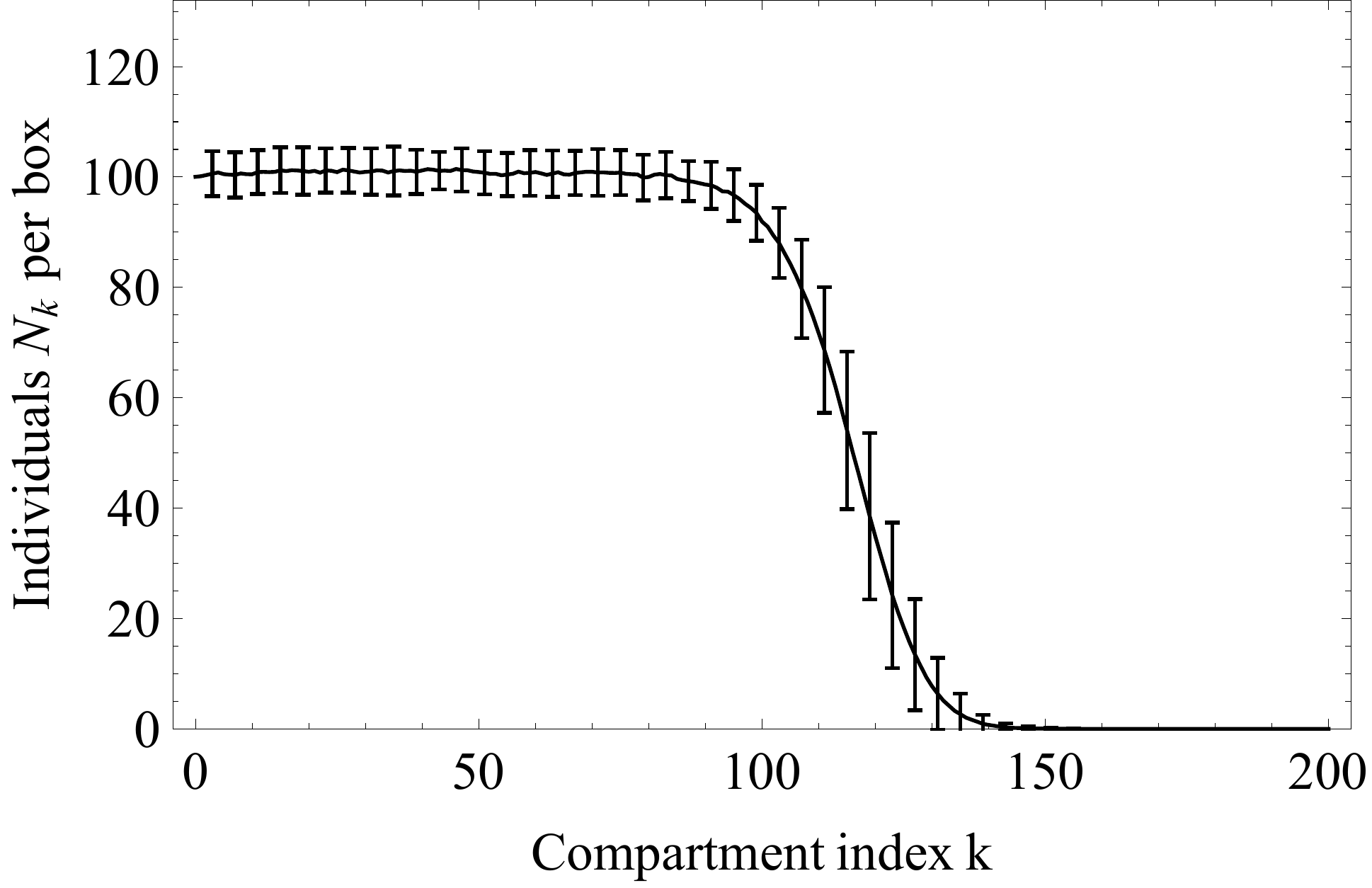}
}
\caption{\changed{Instantaneous travelling wave profile in the stochastic Fisher-Kolmogorov equation} \eqref{eq:MasterEquationFisher} with parameters $D=10,\lambda=1, h=1, k_{max}=200$ and $\Omega=100$. The random walk part of the model used for \protect\subref{fig:FisherTravellingWavesNorm} is based on \eqref{eq:transitionRatesRandomWalk}, whereas for \protect\subref{fig:FisherTravellingWavesOpt} it is based on \eqref{eq:transitionRatesDifference}. The initial condition in each case was a traveling wave approximation and the plots shown were taken after a sufficiently large time to allow a stable travelling wave to form in each case. Each plot shows the average of $256$ different realisations, and the bars indicate the corresponding standard deviation.
\label{fig:FisherTravellingWaves}} 
\end{figure}
\changed{In Figure \ref{fig:FisherTravellingWaves} we present snapshots of the average travelling wave profile from $256$ different realisations of the stochastic Fisher-Kolmogorov equation \eqref{eq:MasterEquationFisher}. In Figure \subref*{fig:FisherTravellingWavesNorm} the random walk is modelled with the exact model, \eqref{eq:transitionRatesRandomWalk}, whereas in Figure \subref*{fig:FisherTravellingWavesOpt} it is based on the difference model, \eqref{eq:transitionRatesDifference}. The results appear indistinguishable, indicating that \eqref{eq:transitionRatesDifference} can reproduce the mean behaviour of \eqref{eq:transitionRatesRandomWalk}. On closer inspection of the standard deviations, shown by the bars, we note that at the wave front (that is, in the regions where the solution involves small, but non-zero $N_k$), Figures \subref*{fig:FisherTravellingWavesNorm} and \subref*{fig:FisherTravellingWavesOpt} look similar, whereas away from the wave front (that is, in the regions where the solution fluctuates around $h\Omega = 100$), \subref*{fig:FisherTravellingWavesOpt} has significantly less noise around the mean than \subref*{fig:FisherTravellingWavesNorm}. The lower noise levels do not appear to have a significant effect on the shape or speed of the travelling wave.}
\begin{figure}[h!]
\includegraphics[width=0.48\linewidth]{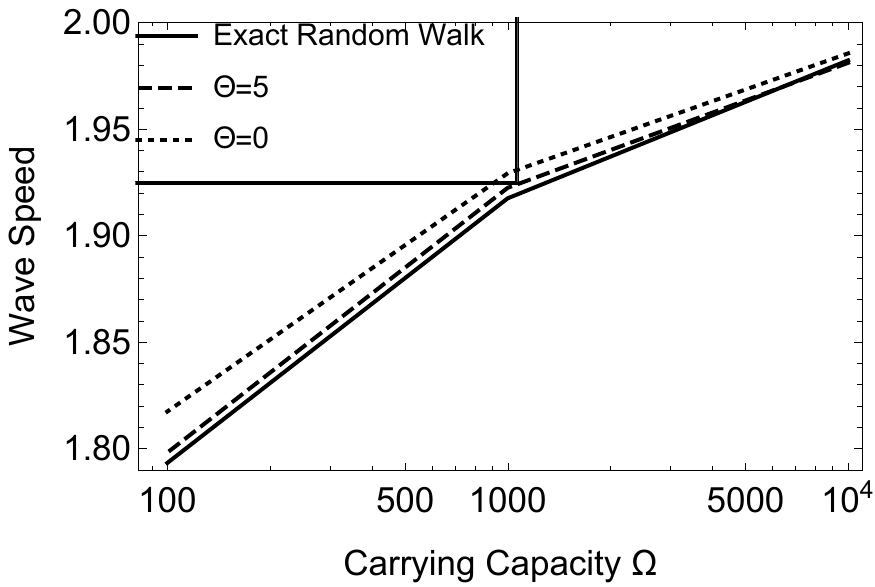}
\caption{Mean travelling wave speed for the stochastic Fisher-Kolmogorov equation \eqref{eq:MasterEquationFisher} when $D=1,\lambda=1, k_{max}=200$ and of the carrying capacity $\Omega$. The solid line shows the wave speed for the conventional random walk model \eqref{eq:transitionRatesRandomWalk}, the dotted line the difference model \eqref{eq:transitionRatesDifference} ($\Theta=0$) and the dashed line the conditional difference model \eqref{eq:transitionRatesMix} ($\Theta=5$). Each plot was obtained by averaging over $256$ realisations and measuring the speed at $5$ time intervals for each simulation.
\label{fig:FisherTravellingWavesSpeed}} 
\end{figure}
\changed{
For a quantitative comparison of the two models, we plot the travelling wave speeds of the exact model \eqref{eq:transitionRatesRandomWalk} as well as the conditional difference model \eqref{eq:transitionRatesMix} in Figure \ref{fig:FisherTravellingWavesSpeed}. This is also a good testing ground for our algoithm, as the wave speed is known to be modified by stochastic effects~\cite{breuer1994fluctuation}, Here, we have obtained the wave speed in the same way as in ~\cite{spill2015hybrid}, averaging over $256$ different simulations in each case, and measuring the speed at $15$ different time intervals. We see that for low carrying capacities ($\Omega=100$), the wave speed obtained by using the random walk difference model \eqref{eq:transitionRatesDifference} ($\Theta=0$) differs by approximately $1\%$ from the travelling wave speed obtained by the conventional random walk \eqref{eq:transitionRatesRandomWalk}, whereas the wave speed from the conditional difference model \eqref{eq:transitionRatesMix} differes by less than $1\%$. Furthermore, increasing the carrying capacity improves the agreement between the models. This is expected, as all three models converge to the mean field model given by equation \eqref{eq:meanEquationFisher} in the limit as $\Omega\to\infty$, and the mean field model has a wave speed close to the continuum model wave speed of $2\sqrt{D\lambda}$ ~\cite{spill2015hybrid}.
}

\begin{figure}[h!]

\subfloat[Exact Model]
{\label{fig:MinSpaceTimeNormal}
\includegraphics[width=0.48\linewidth]{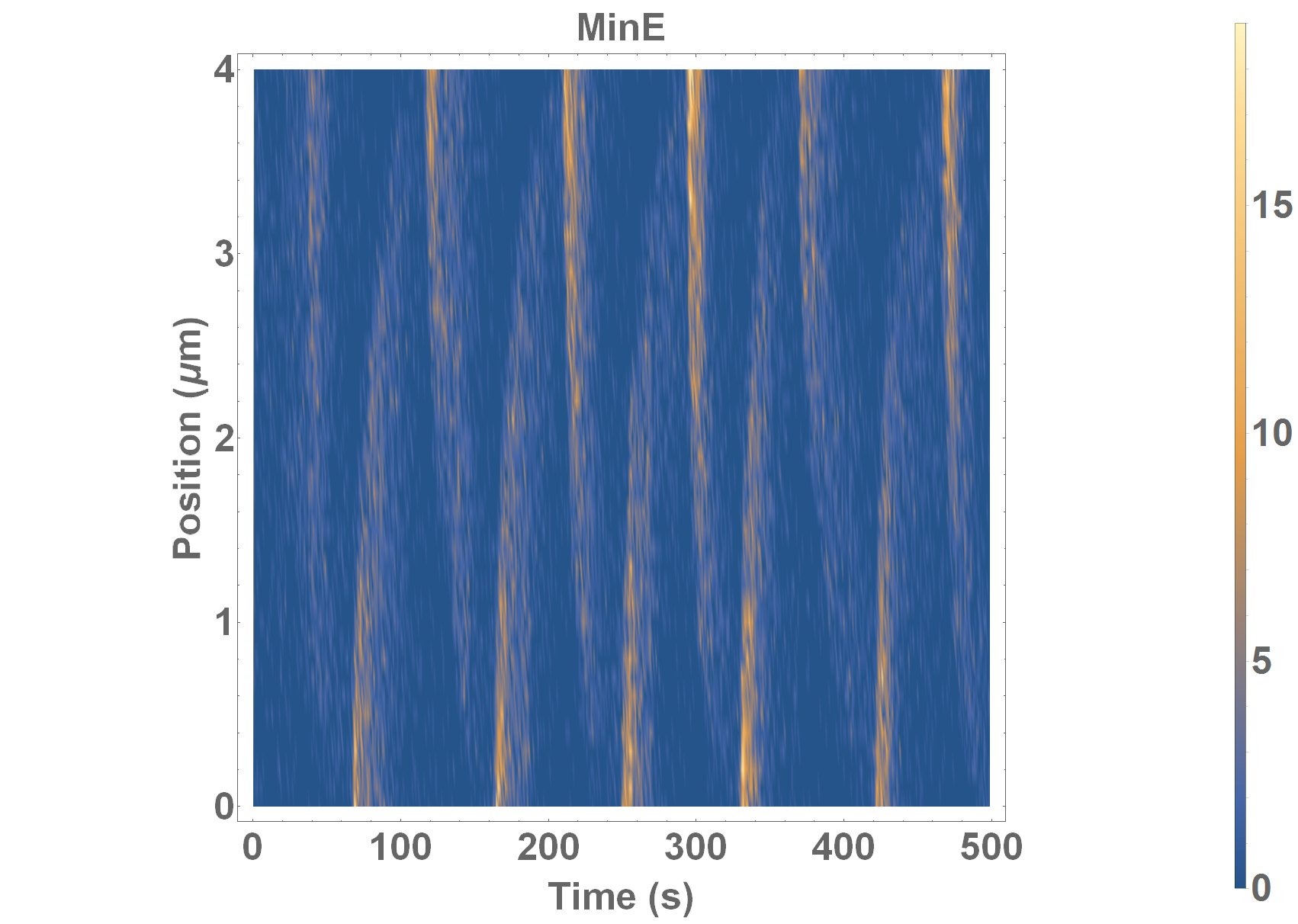}
}
\subfloat[$\Theta = 10$]
{\label{fig:MinSpaceTimeTheta10}
\includegraphics[width=0.48\linewidth]{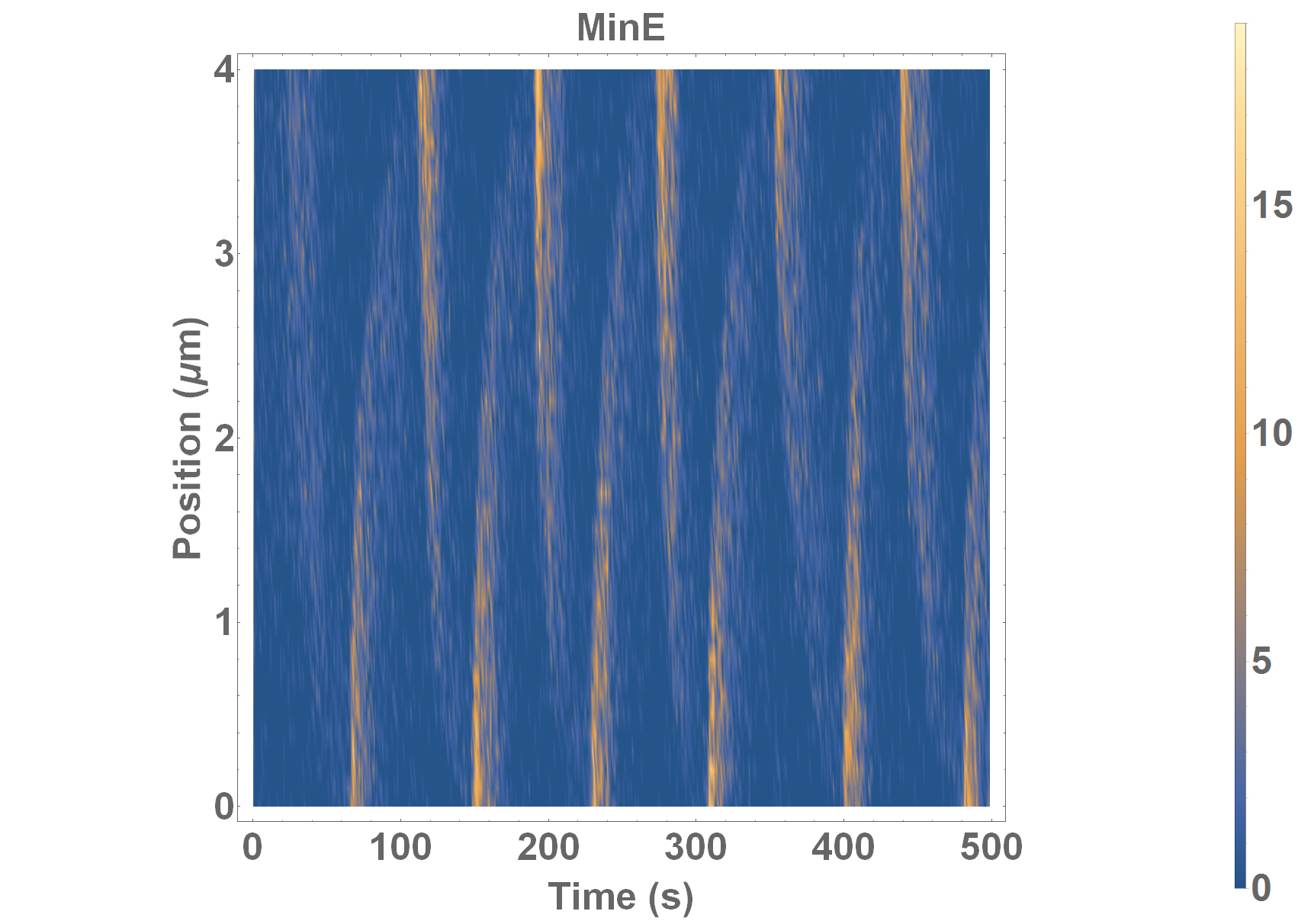}
}\\
\subfloat[$\Theta = 2$]
{\label{fig:MinSpaceTimeTheta2}
\includegraphics[width=0.48\linewidth]{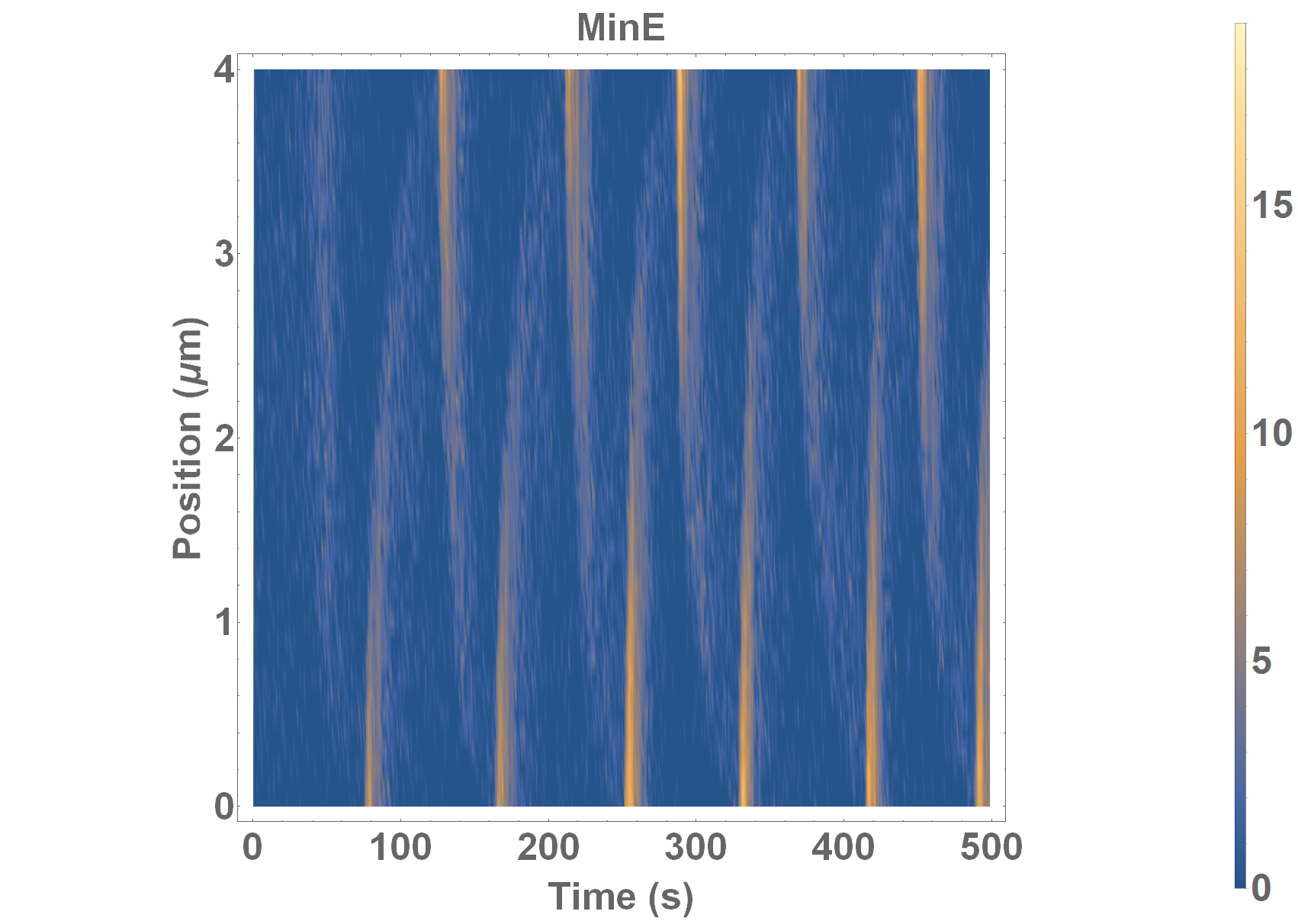}
}
\subfloat[$\Theta = 0$]
{\label{fig:MinSpaceTimeTheta0}
\includegraphics[width=0.48\linewidth]{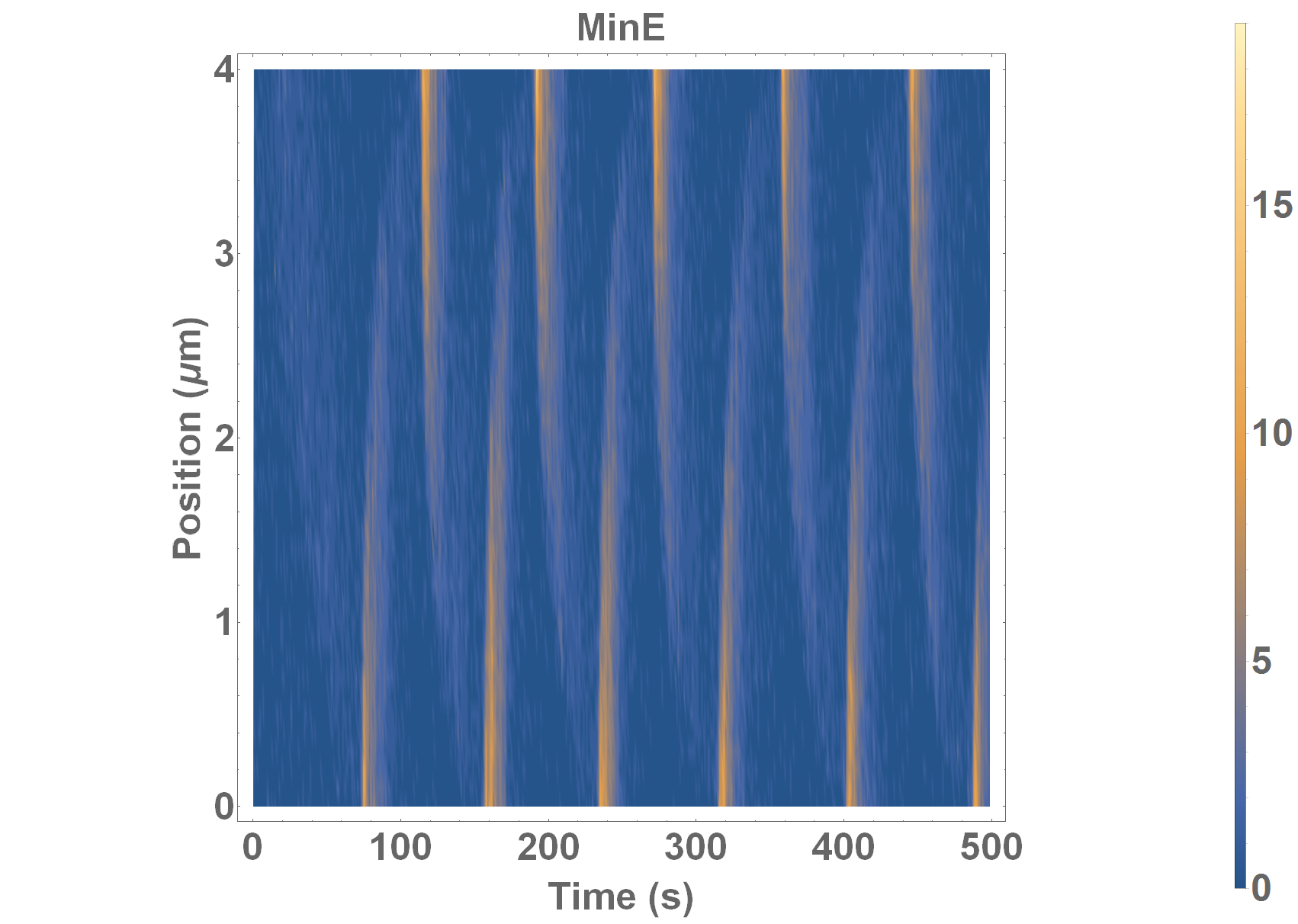}
}
\caption{Space-Time evolution of the number of MinE molecules with an initial number of $1400$ MinE and $6700$ MinD proteins. We  compare simulations of the stochastic model where diffusion is modeled by the hybrid approach, equation \eqref{eq:transitionRatesMix}, with different thresholds. Stable oscillations are forming after less than $100 s$. All four cases appear similar, but the cases with low thresholds, $\Theta =0,2$, appear less noisy than the case $\Theta = 10$ and the case of normal diffusion.
\label{fig:MinSpaceTime}} 
\end{figure}

\changed{
\subsection{Min Oscillations in E.coli}\label{sec:Min}

We now investigate a model describing the interactions of Min proteins in E.coli. The deterministic PDE model was originally described in ~\cite{huang2003dynamic}, and stochastic versions were studied in ~\cite{kerr2006division,fange2006noise}. We focus on a one-dimensional version of this model. The model describes reactions and diffusion of the molecules $MinD$ and $MinE$ which can exist in several states. $MinD^{ADP}$ and $MinD^{ATP}$ denote $MinD$ sequestered in the cytosol and bound to ADP or ATP, respectively, whereas the sequestered form of $MinE$ is simply denoted by $MinE$. The membrane-bound form of $MinD$ is denoted by $MinD(M)$, and $MinDE(M)$ denotes the membrane-bound form of a complex of $MinD$ and $MinE$. 
The reactions are defined by:
\begin{align}
\begin{split}
MinD^{ADP} &\xrightarrow[]{k_1} MinD^{ATP}, \\
MinD^{ATP} &\xrightarrow[]{k_2} MinD(M), \\
MinD^{ATP} + MinD(M) &\xrightarrow[]{k_3} 2 MinD(M), \\
MinD^{ATP} + MinDE(M) &\xrightarrow[]{k_3} MinDE(M) +  MinD(M), \\
MinE + MinD(M) &\xrightarrow[]{k_4} MinDE, \\
MinDE(M) &\xrightarrow[]{k_5} MinD^{ADP}(M) + MinE \\
\end{split}
\end{align}

Furthermore, the cytosolic species $MinD^{ADP}$, $MinD^{ATP}$ and $MinE$ are able to diffuse through the cytosol with a macroscopic diffusion constant $D=2.5 \mu m^2/s$, which translates, as before, into rates on our lattice given by $\frac{D}{h^2}$. We assume diffusion on the membrane of $MinD(M)$ and $MinDE(M)$ is negligible.

 The 1D model is obtained from the 3D model on a cylinder of length $L=4\mu m$ and radius $R=0.5\mu m$, by assuming diffusion occurs only along the main axis. We discretize along the length of the cylinder into cross-sectional disk-shaped compartments of length $h = 0.1\mu m$. Then, the reaction rates of the stochastic reaction, based on the PDE model ~\cite{huang2003dynamic}, are given by $k_1=\sigma_D^{ADP\rightarrow ATP} = 1 s^{-1}$, $k_2 = \frac{2\sigma_D}{R} = 0.1 s^{-1}$, $k_3 = \frac{\sigma_{dD}}{\pi R^2 h} = 0.019 s^{-1}$, $k_4 = \frac{\sigma_E}{\pi R^2 h} = 1.18 s^{-1}$, $k_5 = \sigma_{de} = 0.7 s^{-1}$. Here, we use the original parameters $\sigma$ from, ~\cite{huang2003dynamic}, to estimate the reaction rates. We compare simulations of this system, when diffusion is described by the conditional difference model \eqref{eq:transitionRatesMix}, with others where diffusion is modelled by the exact random walk model \eqref{eq:transitionRatesRandomWalk}.

Supplementary Figure \ref{fig:MinSpaceTime} shows the space-time evolution of the number of cytosolic MinE molecules with an initial total number of $1400$ MinE and $6700$ MinD proteins, for three different threshold values $\Theta = 0,1,10$, and the case of the exact random walk model \eqref{eq:transitionRatesRandomWalk}, which corresponds to the conditional difference model with a large threshold. All four cases yield qualitatively similar results, with stable oscillations appearing shortly before $t=100s$, with a period of approximately $70s$. However, noise levels are higher for the exact model, and when $\Theta=10$, than when $\Theta = 0,1$. We then quantified the distribution of molecular numbers. Figure \ref{fig:Histogram} shows the distribution of four different molecular species, in the center of the simulation domain, corresponding to compartment$k=20$, at time $t=300$ and for $128$ simulations.
\begin{figure}[h!]

\subfloat[]
{\label{fig:HistogramMinDADP}
\includegraphics[width=0.48\linewidth]{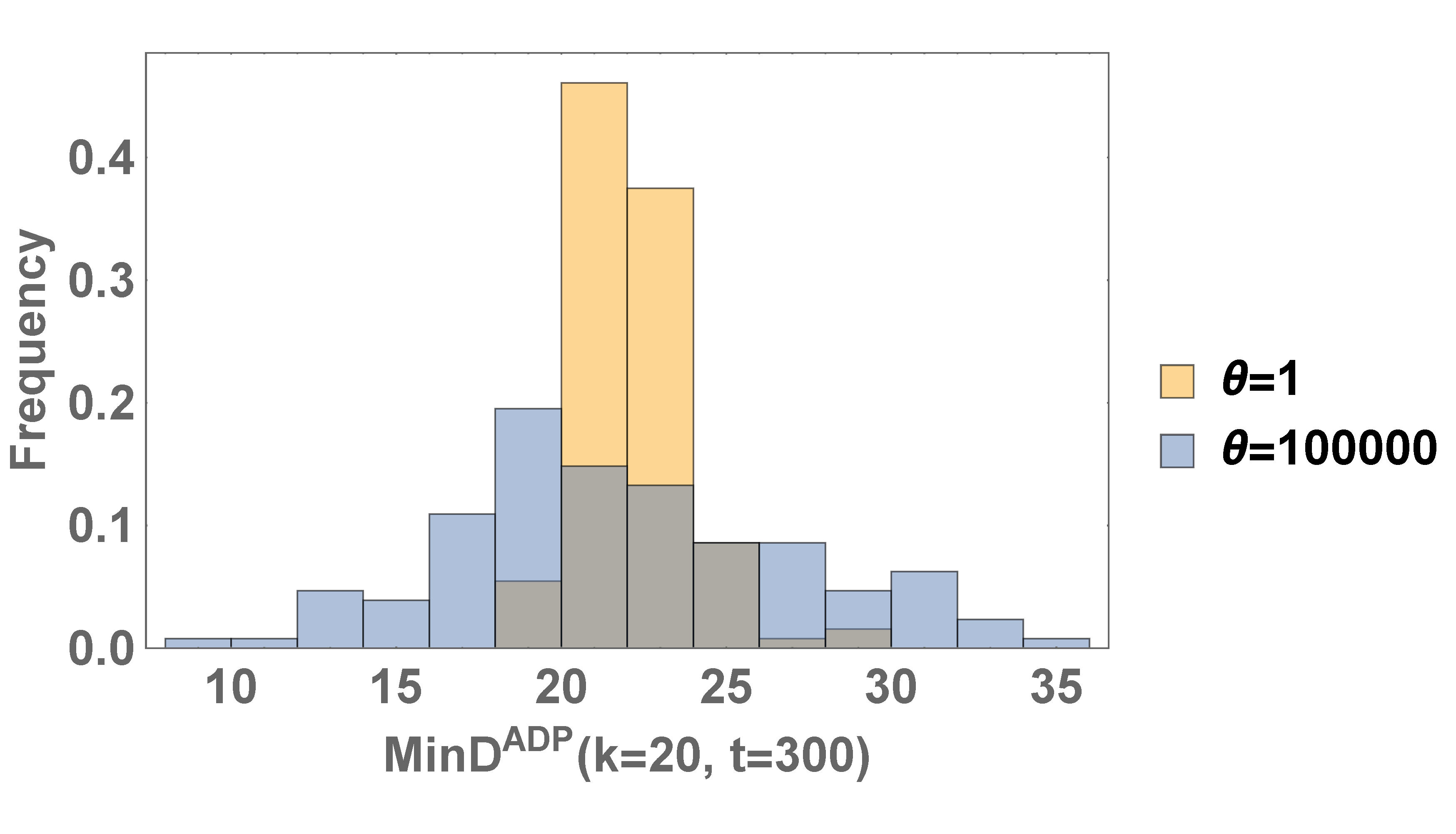}
}
\subfloat[]
{\label{fig:HistogramMinDATP}
\includegraphics[width=0.48\linewidth]{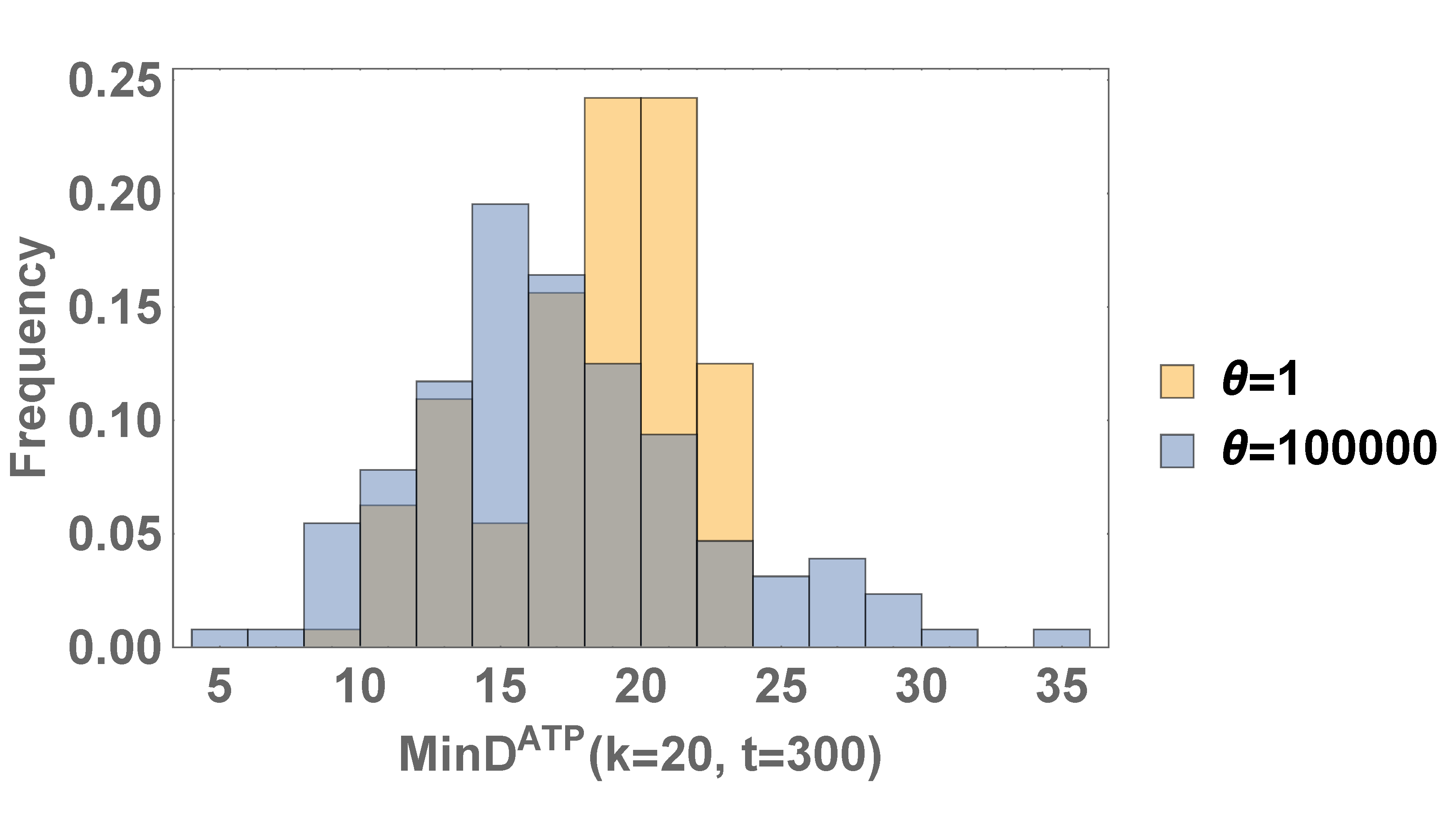}
}\\
\subfloat[]
{\label{fig:HistogramMinE}
\includegraphics[width=0.48\linewidth]{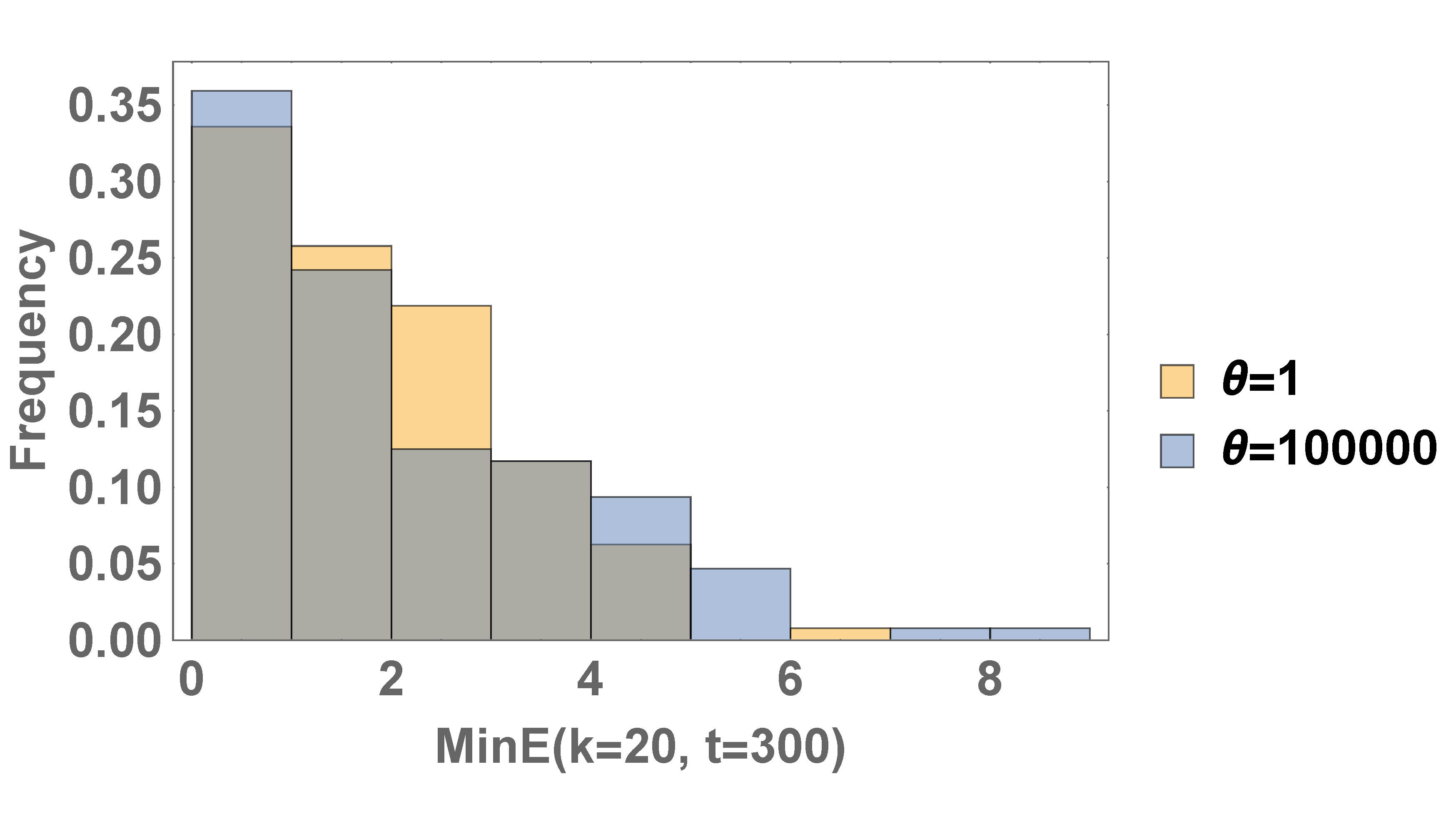}
}
\subfloat[]
{\label{fig:HistogramMinDM}
\includegraphics[width=0.48\linewidth]{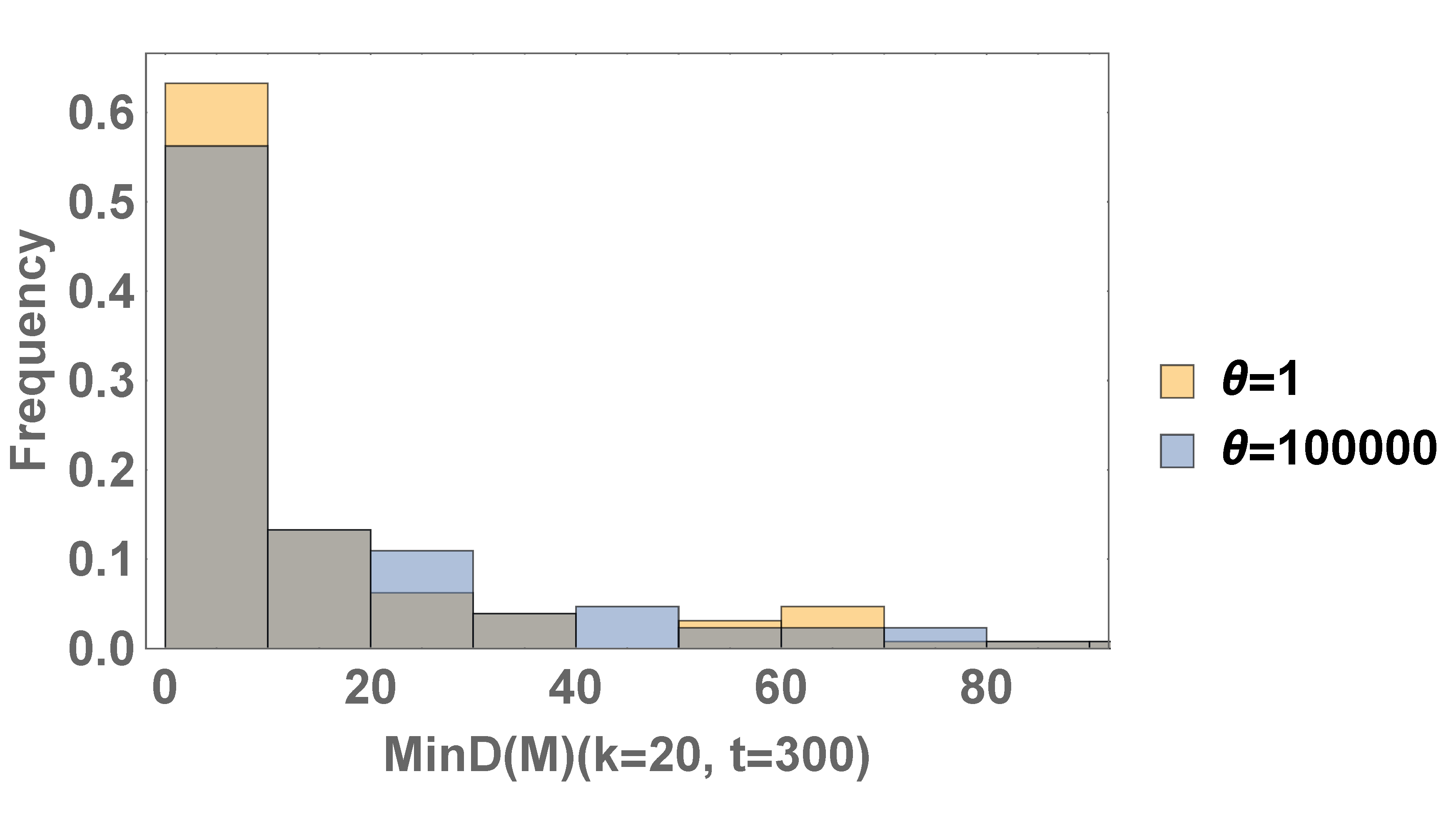}
}
\caption{Distribution of the number of Min molecules with an initial number of $1400$ MinE and $6700$ MinD proteins. We  compare simulations of the stochastic model where diffusion is modeled by the conditional difference model, equation \eqref{eq:transitionRatesMix}, with a threshold of $\Theta = 1$, and the case of the exact model, which is here obtained by choosing a threshold of $\Theta = 100000$. The histograms are obtained from the molecular numbers in a central compartment, $k=20$, at time $t=300s$, from $128$ simulations of the model.
\label{fig:Histogram}} 
\end{figure}
We see that $MinD^{ADP}$ and $MinD^{ATP}$ numbers (Figures \subref*{fig:HistogramMinDADP}, \subref*{fig:HistogramMinDATP}) are unimodally distributed around a peak, and the conditional difference model with $\Theta=1$ yields a narrower distribution than the exact model. By contrast, the distributions for $MinE$ and $MinD(M)$ (Figures \subref*{fig:HistogramMinE}, \subref*{fig:HistogramMinDM}) appear to decay monotonically with a peak at $0$, and we observe no significant deviations between the exact model and the conditional difference model. To obtain more significant data, we average the standard deviations of those distributions over $200$ time measurements (from $t=300s$ to $t=500s$, with measurements every $1s$ to avoid transient effects associated with the formation of the oscillations), in each compartment, and from each of the $128$ different simulations. 
\begin{figure}[h!]

\subfloat[]
{\label{fig:MinStandardDeviationsMinDADP}
\includegraphics[width=0.48\linewidth]{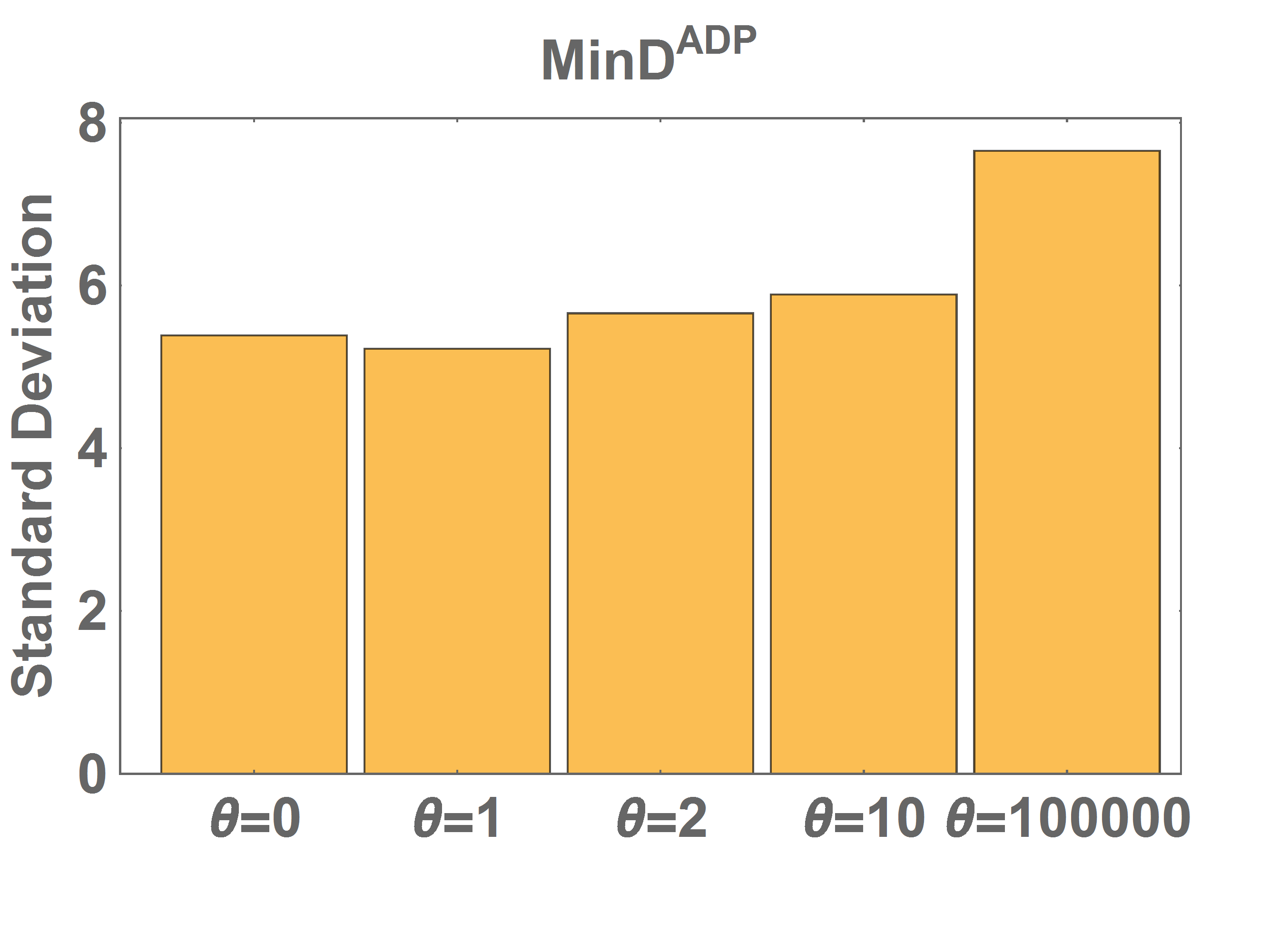}
}
\subfloat[]
{\label{fig:MinStandardDeviationsMinDATP}
\includegraphics[width=0.48\linewidth]{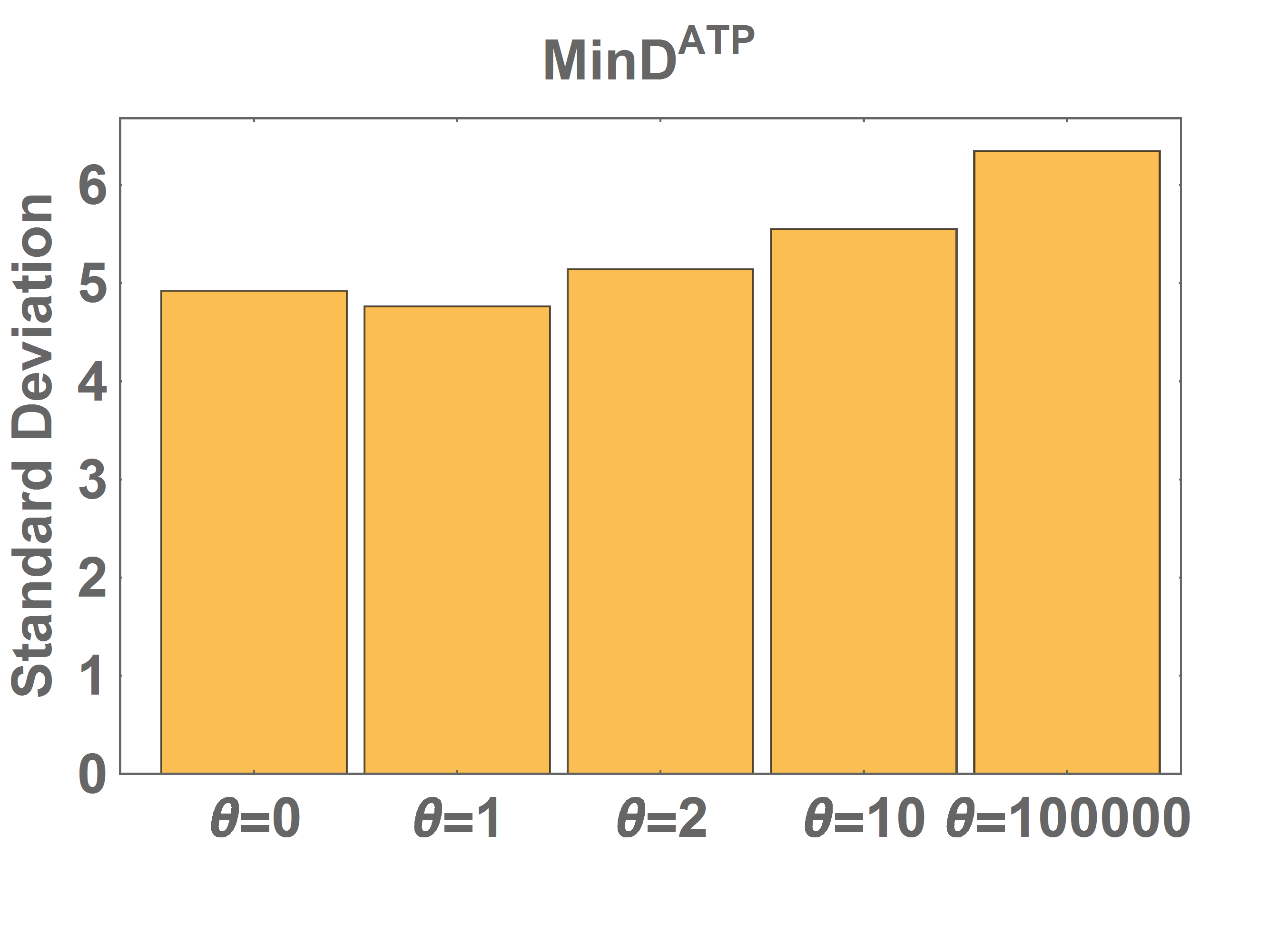}
}\\
\subfloat[]
{\label{fig:MinStandardDeviationsMinE}
\includegraphics[width=0.48\linewidth]{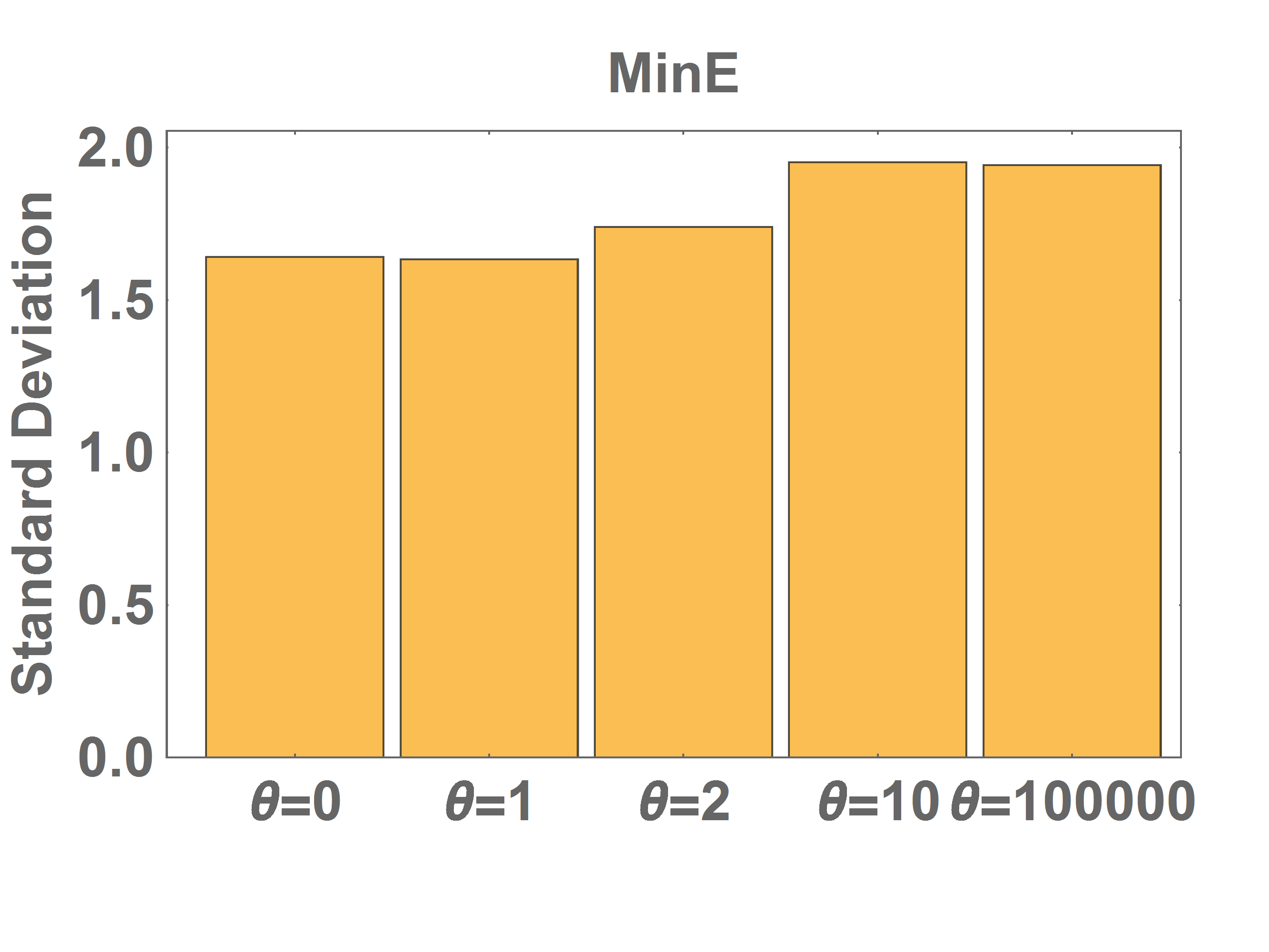}
}
\subfloat[]
{\label{fig:MinStandardDeviationsMinDM}
\includegraphics[width=0.48\linewidth]{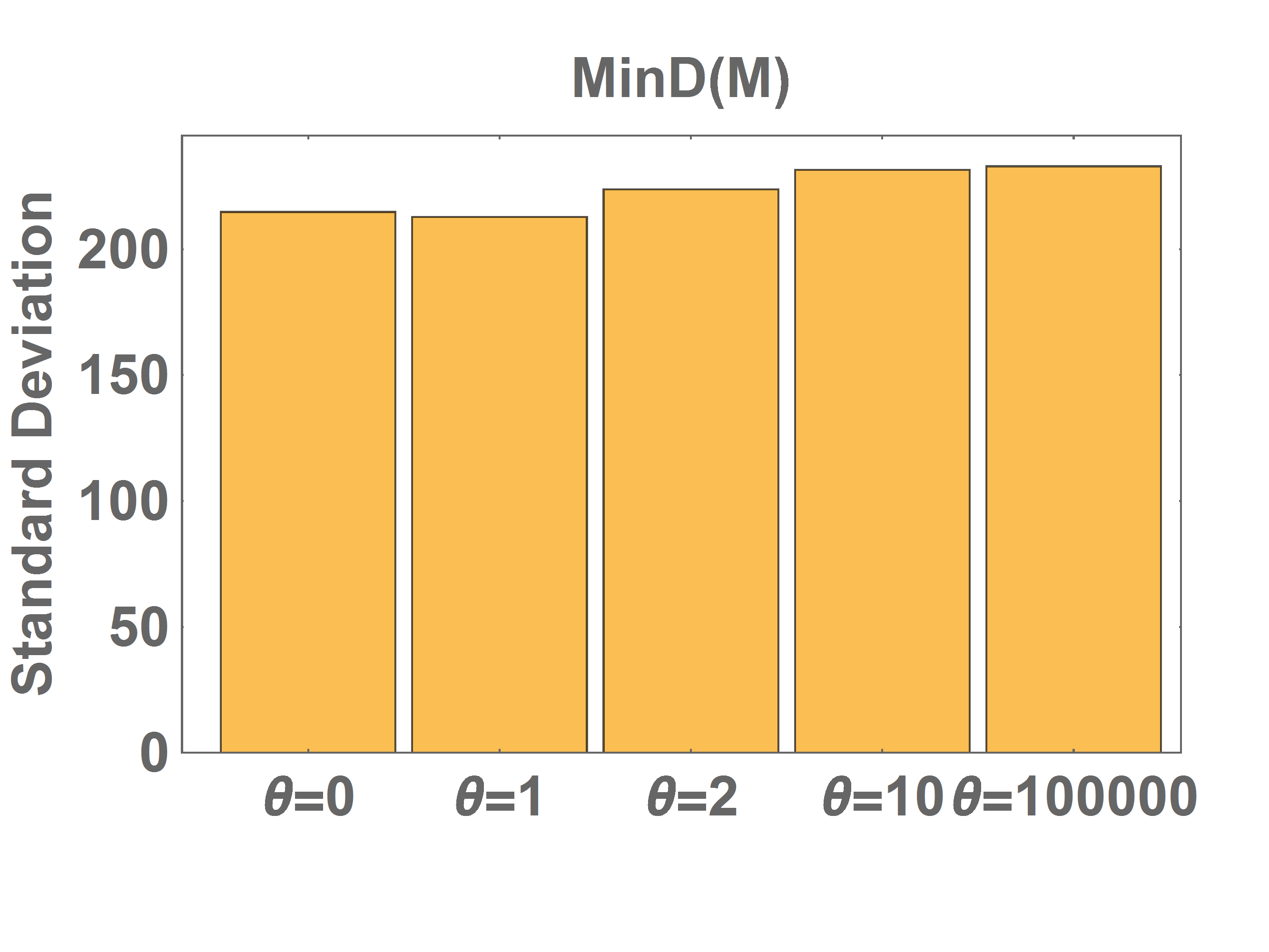}
}
\caption{Standard deviations from the distributions shown in \ref{fig:Histogram}. We have performed $128$ simulations, and averaged the standard deviations over each space and time point, from $t=300s$ until $t=500s$. 
\label{fig:MinStandardDeviations}} 
\end{figure}
Figure \ref{fig:MinStandardDeviations} shows these standard deviations for five threshold values ($\Theta = 0,1,2,10,100000$), where the last case coincides with the exact model. We see that for $MinE$ and $MinD(M)$, there is excellent agreement between the different thresholds, whereas the standard deviations of the conditional difference models are slightly lower than those for the exact model for $MinD^{ADP}$ and $MinD^{ATP}$. However, in those cases, relative noise is small as particle numbers are, on average, relatively large (see supplementary Figure \ref{fig:MinMean}, which also confirms that the means are almost identical for all chosen thresholds). 
Focusing on performance, we note that the computational cost decreases dramatically as $\Theta$ decreases.	
\begin{figure}[h!]
\includegraphics[width=0.48\linewidth]{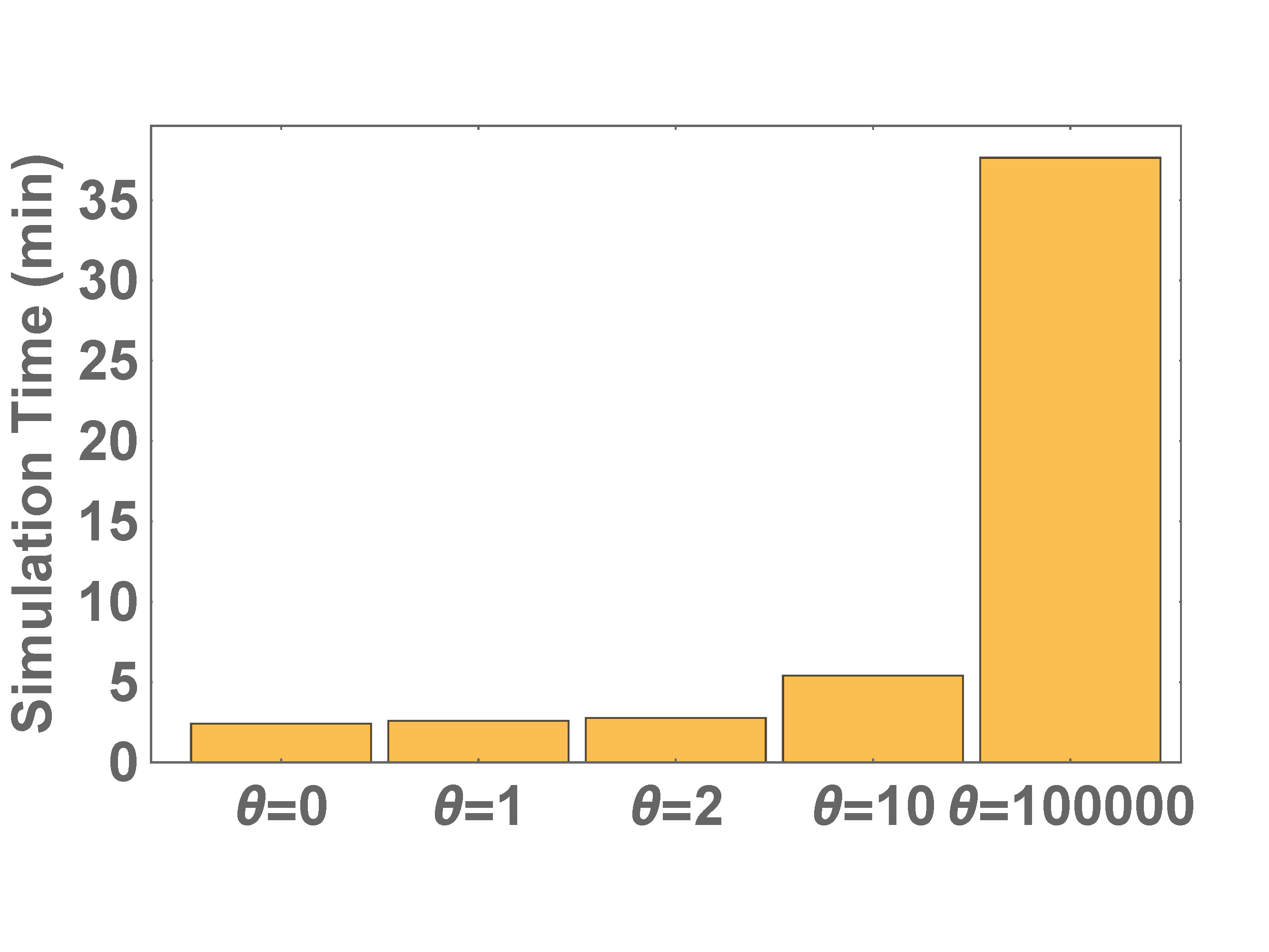}
\caption{Simulation time on a $3$ GZHz Xeon machine of the simulations shown in Figures \ref{fig:Histogram} and \ref{fig:MinStandardDeviations}.
\label{fig:MinPerformance}} 
\end{figure}
Figure $\ref{fig:MinPerformance}$ shows that absolute simulation time when $\Theta=10$ decreased about $7$ compared to the exact model, with a further $2$ fold gain for thresholds of $\Theta=0,1,2$. Based on these results, we conclude that choosing a threshold of $\Theta=10$ offers excellent performance gains while maintaining the stochastic fluctuations associated with the exact model.

We stress that performance gains and quantitative similarity are parameter dependent. In supplementary  Figure \ref{fig:MinStandardDeviations_350}, we present results for the same scenario as before, but with only $25\%$ of the particles present. We note that the standard deviations, similar to those shown in Figure \ref{fig:MinStandardDeviations}, are similar for different thresholds. Corresponding performance gains are shown in Figure \ref{fig:MinPerformance_350}. The gains are small for $\Theta = 10$, but for $\Theta = 0,1,2$ the performance gains range from $13$ to $25$ fold. Hence, in this case, fixing $\Theta=2$ may be preferred, as it offers great performance gains, but the behaviour of the species with the highest levels of relative fluctuation, $MinE$, behaves in a similar manner to the case of the exact random walk model.

}

%
%
\section{Conclusions}
\changed{In this paper, we have presented an algorithm which alters the transition rates that appear in reactions of the chemical master equation so that when two reactions have opposite effects on the state of the system, and when the particle number exceeds a threshold value, we replace the two reactions by a single reaction which represents the net effect of these two reactions.} The mean behaviour of the stochastic model is preserved by this change of transition rates, and by setting the threshold sufficiently high it is possible to preserve essential stochastic features of the original model, such as variances, first-passage time in random walk models, or the stochastically modified wave speed in the stochastic Fisher-Kolmogorov equation. 

We have applied the algorithm to diffusion and reaction-diffusion problems as diffusion often generates many practically unobservable events such as the swapping of positions of two identical particles, and such events are easily eliminated by our algorithm. Furthermore, as diffusion rates can often be considerably higher than reaction rates our algorithm can help to significantly speed up simulations of the stochastic reaction-diffusion systems while preserving essential stochastic features. \changed{We have confirmed this for the problem of first-passage time, the modification of the wave speed of the Fisher-Kolmogorov system due to stochastic effects, and the distribution of particles in low copy numbers in the oscillatory system of Min proteins. Our current focus was on one-dimensional models. Future work will include a more detailed investigation of lattice dependencies, which can have an important effect on the behaviour of stochastic reaction-diffusion systems ~\cite{isaacson2013convergent,gillespie2014validity,hellander2015reaction}, or generalizations to more complex geometries as carried out in~\cite{isaacson2006incorporating,drawert2012urdme}. There is also a need to test our method on simple lattices in higher dimensions. While the number of Gillespie events will be reduced in a similar manner as in one spatial dimension, events in the conditional difference model will lead to shifts in the total propensity, whereas movements in the bulk of the exact random walk model will only lead to local changes of propensities. This might make optimisations of the Gillespie algorithm in terms of efficient calculation of propensities potentially difficult.}

A main advantage of our method for increasing the simulation speed of stochastic reaction or reaction-diffusion models is its simplicity of implementation. Unlike other algorithms designed to improve the simulation speed of such models, such as ~\cite{gibson2000,moro2004,hellander2007hybrid,henzinger2010hybrid,spill2015hybrid,gillespie2001approximate,gillespie2007stochastic,cao2006efficient,gillespie2003improved}, ours can be implemented by simply redefining the transition rates in the Gillespie algorithm. \changed{Furthermore, it does not depend on the actual implementation used to simulate the stochastic process. Whilst we have used the Gillespie algorithm in this paper, it is straightforward to combine our algorithm with other solution methods of the master equation including the algorithms mentioned above.}

\begin{acknowledgments}
This publication was based on work supported in part by Award No KUK-C1-013-04, made by King Abdullah University of Science and Technology (KAUST). We are grateful to R. Erban, M. Flegg, A. McKane and M. Robinson for helpful discussions, and the anonymous referees for their helpful suggestions.
\end{acknowledgments}

%


\begin{thebibliography}{42}%
\makeatletter
\providecommand \@ifxundefined [1]{%
 \@ifx{#1\undefined}
}%
\providecommand \@ifnum [1]{%
 \ifnum #1\expandafter \@firstoftwo
 \else \expandafter \@secondoftwo
 \fi
}%
\providecommand \@ifx [1]{%
 \ifx #1\expandafter \@firstoftwo
 \else \expandafter \@secondoftwo
 \fi
}%
\providecommand \natexlab [1]{#1}%
\providecommand \enquote  [1]{``#1''}%
\providecommand \bibnamefont  [1]{#1}%
\providecommand \bibfnamefont [1]{#1}%
\providecommand \citenamefont [1]{#1}%
\providecommand \href@noop [0]{\@secondoftwo}%
\providecommand \href [0]{\begingroup \@sanitize@url \@href}%
\providecommand \@href[1]{\@@startlink{#1}\@@href}%
\providecommand \@@href[1]{\endgroup#1\@@endlink}%
\providecommand \@sanitize@url [0]{\catcode `\\12\catcode `\$12\catcode
  `\&12\catcode `\#12\catcode `\^12\catcode `\_12\catcode `\%12\relax}%
\providecommand \@@startlink[1]{}%
\providecommand \@@endlink[0]{}%
\providecommand \url  [0]{\begingroup\@sanitize@url \@url }%
\providecommand \@url [1]{\endgroup\@href {#1}{\urlprefix }}%
\providecommand \urlprefix  [0]{URL }%
\providecommand \Eprint [0]{\href }%
\providecommand \doibase [0]{http://dx.doi.org/}%
\providecommand \selectlanguage [0]{\@gobble}%
\providecommand \bibinfo  [0]{\@secondoftwo}%
\providecommand \bibfield  [0]{\@secondoftwo}%
\providecommand \translation [1]{[#1]}%
\providecommand \BibitemOpen [0]{}%
\providecommand \bibitemStop [0]{}%
\providecommand \bibitemNoStop [0]{.\EOS\space}%
\providecommand \EOS [0]{\spacefactor3000\relax}%
\providecommand \BibitemShut  [1]{\csname bibitem#1\endcsname}%
\let\auto@bib@innerbib\@empty
\bibitem [{\citenamefont {Van~Kampen}(1992)}]{van1992stochastic}%
  \BibitemOpen
  \bibfield  {author} {\bibinfo {author} {\bibfnamefont {N.~G.}\ \bibnamefont
  {Van~Kampen}},\ }\href@noop {} {\emph {\bibinfo {title} {Stochastic Processes
  in Physics and Chemistry}}},\ Vol.~\bibinfo {volume} {1}\ (\bibinfo
  {publisher} {North holland},\ \bibinfo {year} {1992})\BibitemShut {NoStop}%
\bibitem [{\citenamefont {Gardiner}(2010)}]{gardiner2010stochastic}%
  \BibitemOpen
  \bibfield  {author} {\bibinfo {author} {\bibfnamefont {C.}~\bibnamefont
  {Gardiner}},\ }\href@noop {} {\emph {\bibinfo {title} {Stochastic Methods}}}\
  (\bibinfo  {publisher} {Springer},\ \bibinfo {year} {2010})\BibitemShut
  {NoStop}%
\bibitem [{\citenamefont {Gillespie}, \citenamefont {Hellander},\ and\
  \citenamefont {Petzold}(2013)}]{gillespie2013perspective}%
  \BibitemOpen
  \bibfield  {author} {\bibinfo {author} {\bibfnamefont {D.~T.}\ \bibnamefont
  {Gillespie}}, \bibinfo {author} {\bibfnamefont {A.}~\bibnamefont
  {Hellander}}, \ and\ \bibinfo {author} {\bibfnamefont {L.~R.}\ \bibnamefont
  {Petzold}},\ }\href@noop {} {\bibfield  {journal} {\bibinfo  {journal} {The
  Journal of Chemical Physics}\ }\textbf {\bibinfo {volume} {138}},\ \bibinfo
  {pages} {170901} (\bibinfo {year} {2013})}\BibitemShut {NoStop}%
\bibitem [{\citenamefont {de~la Cruz}\ \emph {et~al.}(2015)\citenamefont {de~la
  Cruz}, \citenamefont {Guerrero}, \citenamefont {Spill},\ and\ \citenamefont
  {Alarc{\'o}n}}]{de2015effects}%
  \BibitemOpen
  \bibfield  {author} {\bibinfo {author} {\bibfnamefont {R.}~\bibnamefont
  {de~la Cruz}}, \bibinfo {author} {\bibfnamefont {P.}~\bibnamefont
  {Guerrero}}, \bibinfo {author} {\bibfnamefont {F.}~\bibnamefont {Spill}}, \
  and\ \bibinfo {author} {\bibfnamefont {T.}~\bibnamefont {Alarc{\'o}n}},\
  }\href@noop {} {\bibfield  {journal} {\bibinfo  {journal} {The Journal of
  Chemical Physics}\ }\textbf {\bibinfo {volume} {143}},\ \bibinfo {pages}
  {074105} (\bibinfo {year} {2015})}\BibitemShut {NoStop}%
\bibitem [{\citenamefont {Dobramysl}, \citenamefont {R{\"u}diger},\ and\
  \citenamefont {Erban}(2015)}]{dobramysl2015particle}%
  \BibitemOpen
  \bibfield  {author} {\bibinfo {author} {\bibfnamefont {U.}~\bibnamefont
  {Dobramysl}}, \bibinfo {author} {\bibfnamefont {S.}~\bibnamefont
  {R{\"u}diger}}, \ and\ \bibinfo {author} {\bibfnamefont {R.}~\bibnamefont
  {Erban}},\ }\href@noop {} {\bibfield  {journal} {\bibinfo  {journal} {arXiv
  preprint arXiv:1504.00146}\ } (\bibinfo {year} {2015})}\BibitemShut {NoStop}%
\bibitem [{\citenamefont {Black}\ and\ \citenamefont
  {McKane}(2012)}]{black2012stochastic}%
  \BibitemOpen
  \bibfield  {author} {\bibinfo {author} {\bibfnamefont {A.~J.}\ \bibnamefont
  {Black}}\ and\ \bibinfo {author} {\bibfnamefont {A.~J.}\ \bibnamefont
  {McKane}},\ }\href@noop {} {\bibfield  {journal} {\bibinfo  {journal} {Trends
  in Ecology \& Evolution}\ }\textbf {\bibinfo {volume} {27}},\ \bibinfo
  {pages} {337} (\bibinfo {year} {2012})}\BibitemShut {NoStop}%
\bibitem [{\citenamefont {Spill}\ \emph {et~al.}(2014)\citenamefont {Spill},
  \citenamefont {Guerrero}, \citenamefont {Alarcon}, \citenamefont {Maini},\
  and\ \citenamefont {Byrne}}]{spill2014mesoscopic}%
  \BibitemOpen
  \bibfield  {author} {\bibinfo {author} {\bibfnamefont {F.}~\bibnamefont
  {Spill}}, \bibinfo {author} {\bibfnamefont {P.}~\bibnamefont {Guerrero}},
  \bibinfo {author} {\bibfnamefont {T.}~\bibnamefont {Alarcon}}, \bibinfo
  {author} {\bibfnamefont {P.~K.}\ \bibnamefont {Maini}}, \ and\ \bibinfo
  {author} {\bibfnamefont {H.~M.}\ \bibnamefont {Byrne}},\ }\href@noop {}
  {\bibfield  {journal} {\bibinfo  {journal} {Journal of Mathematical Biology}\
  }\textbf {\bibinfo {volume} {70}},\ \bibinfo {pages} {485} (\bibinfo {year}
  {2014})}\BibitemShut {NoStop}%
\bibitem [{\citenamefont {Sturrock}\ \emph {et~al.}(2014)\citenamefont
  {Sturrock}, \citenamefont {Hellander}, \citenamefont {Aldakheel},
  \citenamefont {Petzold},\ and\ \citenamefont {Chaplain}}]{sturrock2014role}%
  \BibitemOpen
  \bibfield  {author} {\bibinfo {author} {\bibfnamefont {M.}~\bibnamefont
  {Sturrock}}, \bibinfo {author} {\bibfnamefont {A.}~\bibnamefont {Hellander}},
  \bibinfo {author} {\bibfnamefont {S.}~\bibnamefont {Aldakheel}}, \bibinfo
  {author} {\bibfnamefont {L.}~\bibnamefont {Petzold}}, \ and\ \bibinfo
  {author} {\bibfnamefont {M.~A.}\ \bibnamefont {Chaplain}},\ }\href@noop {}
  {\bibfield  {journal} {\bibinfo  {journal} {Bulletin of Mathematical
  Biology}\ }\textbf {\bibinfo {volume} {76}},\ \bibinfo {pages} {766}
  (\bibinfo {year} {2014})}\BibitemShut {NoStop}%
\bibitem [{\citenamefont {Guerrero}\ \emph {et~al.}(2015)\citenamefont
  {Guerrero}, \citenamefont {Byrne}, \citenamefont {Maini},\ and\ \citenamefont
  {Alarc{\'o}n}}]{guerrero2015invasion}%
  \BibitemOpen
  \bibfield  {author} {\bibinfo {author} {\bibfnamefont {P.}~\bibnamefont
  {Guerrero}}, \bibinfo {author} {\bibfnamefont {H.~M.}\ \bibnamefont {Byrne}},
  \bibinfo {author} {\bibfnamefont {P.~K.}\ \bibnamefont {Maini}}, \ and\
  \bibinfo {author} {\bibfnamefont {T.}~\bibnamefont {Alarc{\'o}n}},\
  }\href@noop {} {\bibfield  {journal} {\bibinfo  {journal} {Journal of
  Mathematical Biology}\ ,\ \bibinfo {pages} {1}} (\bibinfo {year}
  {2015})}\BibitemShut {NoStop}%
\bibitem [{\citenamefont {Gillespie}(1976)}]{gillespie1976general}%
  \BibitemOpen
  \bibfield  {author} {\bibinfo {author} {\bibfnamefont {D.~T.}\ \bibnamefont
  {Gillespie}},\ }\href@noop {} {\bibfield  {journal} {\bibinfo  {journal}
  {Journal of Computational Physics}\ }\textbf {\bibinfo {volume} {22}},\
  \bibinfo {pages} {403} (\bibinfo {year} {1976})}\BibitemShut {NoStop}%
\bibitem [{\citenamefont {Gillespie}(1977)}]{gillespie1977exact}%
  \BibitemOpen
  \bibfield  {author} {\bibinfo {author} {\bibfnamefont {D.~T.}\ \bibnamefont
  {Gillespie}},\ }\href@noop {} {\bibfield  {journal} {\bibinfo  {journal} {The
  Journal of Physical Chemistry}\ }\textbf {\bibinfo {volume} {81}},\ \bibinfo
  {pages} {2340} (\bibinfo {year} {1977})}\BibitemShut {NoStop}%
\bibitem [{\citenamefont {Gibson}\ and\ \citenamefont
  {Bruck}(2000)}]{gibson2000}%
  \BibitemOpen
  \bibfield  {author} {\bibinfo {author} {\bibfnamefont {M.~A.}\ \bibnamefont
  {Gibson}}\ and\ \bibinfo {author} {\bibfnamefont {J.}~\bibnamefont {Bruck}},\
  }\href@noop {} {\bibfield  {journal} {\bibinfo  {journal} {J. Phys. Chem. A}\
  }\textbf {\bibinfo {volume} {104}},\ \bibinfo {pages} {1876} (\bibinfo {year}
  {2000})}\BibitemShut {NoStop}%
\bibitem [{\citenamefont {Breuer}, \citenamefont {Huber},\ and\ \citenamefont
  {Petruccione}(1995)}]{moro2004}%
  \BibitemOpen
  \bibfield  {author} {\bibinfo {author} {\bibfnamefont {H.-P.}\ \bibnamefont
  {Breuer}}, \bibinfo {author} {\bibfnamefont {W.}~\bibnamefont {Huber}}, \
  and\ \bibinfo {author} {\bibfnamefont {F.}~\bibnamefont {Petruccione}},\
  }\href@noop {} {\bibfield  {journal} {\bibinfo  {journal} {EPL (Europhysics
  Letters)}\ }\textbf {\bibinfo {volume} {30}},\ \bibinfo {pages} {69}
  (\bibinfo {year} {1995})}\BibitemShut {NoStop}%
\bibitem [{\citenamefont {Hellander}\ and\ \citenamefont
  {L{\"o}tstedt}(2007)}]{hellander2007hybrid}%
  \BibitemOpen
  \bibfield  {author} {\bibinfo {author} {\bibfnamefont {A.}~\bibnamefont
  {Hellander}}\ and\ \bibinfo {author} {\bibfnamefont {P.}~\bibnamefont
  {L{\"o}tstedt}},\ }\href@noop {} {\bibfield  {journal} {\bibinfo  {journal}
  {Journal of Computational Physics}\ }\textbf {\bibinfo {volume} {227}},\
  \bibinfo {pages} {100} (\bibinfo {year} {2007})}\BibitemShut {NoStop}%
\bibitem [{\citenamefont {Henzinger}\ \emph {et~al.}(2010)\citenamefont
  {Henzinger}, \citenamefont {Mikeev}, \citenamefont {Mateescu},\ and\
  \citenamefont {Wolf}}]{henzinger2010hybrid}%
  \BibitemOpen
  \bibfield  {author} {\bibinfo {author} {\bibfnamefont {T.~A.}\ \bibnamefont
  {Henzinger}}, \bibinfo {author} {\bibfnamefont {L.}~\bibnamefont {Mikeev}},
  \bibinfo {author} {\bibfnamefont {M.}~\bibnamefont {Mateescu}}, \ and\
  \bibinfo {author} {\bibfnamefont {V.}~\bibnamefont {Wolf}},\ }in\ \href@noop
  {} {\emph {\bibinfo {booktitle} {Proceedings of the 8th International
  Conference on Computational Methods in Systems Biology}}}\ (\bibinfo
  {organization} {ACM},\ \bibinfo {year} {2010})\ pp.\ \bibinfo {pages}
  {55--65}\BibitemShut {NoStop}%
\bibitem [{\citenamefont {Spill}\ \emph {et~al.}(2015)\citenamefont {Spill},
  \citenamefont {Guerrero}, \citenamefont {Alarcon}, \citenamefont {Maini},\
  and\ \citenamefont {Byrne}}]{spill2015hybrid}%
  \BibitemOpen
  \bibfield  {author} {\bibinfo {author} {\bibfnamefont {F.}~\bibnamefont
  {Spill}}, \bibinfo {author} {\bibfnamefont {P.}~\bibnamefont {Guerrero}},
  \bibinfo {author} {\bibfnamefont {T.}~\bibnamefont {Alarcon}}, \bibinfo
  {author} {\bibfnamefont {P.~K.}\ \bibnamefont {Maini}}, \ and\ \bibinfo
  {author} {\bibfnamefont {H.}~\bibnamefont {Byrne}},\ }\href@noop {}
  {\bibfield  {journal} {\bibinfo  {journal} {Journal of Computational
  Physics}\ }\textbf {\bibinfo {volume} {299}},\ \bibinfo {pages} {429}
  (\bibinfo {year} {2015})}\BibitemShut {NoStop}%
\bibitem [{\citenamefont {Gillespie}(2001)}]{gillespie2001approximate}%
  \BibitemOpen
  \bibfield  {author} {\bibinfo {author} {\bibfnamefont {D.~T.}\ \bibnamefont
  {Gillespie}},\ }\href@noop {} {\bibfield  {journal} {\bibinfo  {journal} {The
  Journal of Chemical Physics}\ }\textbf {\bibinfo {volume} {115}},\ \bibinfo
  {pages} {1716} (\bibinfo {year} {2001})}\BibitemShut {NoStop}%
\bibitem [{\citenamefont {Gillespie}\ and\ \citenamefont
  {Petzold}(2003)}]{gillespie2003improved}%
  \BibitemOpen
  \bibfield  {author} {\bibinfo {author} {\bibfnamefont {D.~T.}\ \bibnamefont
  {Gillespie}}\ and\ \bibinfo {author} {\bibfnamefont {L.~R.}\ \bibnamefont
  {Petzold}},\ }\href@noop {} {\bibfield  {journal} {\bibinfo  {journal} {The
  Journal of Chemical Physics}\ }\textbf {\bibinfo {volume} {119}},\ \bibinfo
  {pages} {8229} (\bibinfo {year} {2003})}\BibitemShut {NoStop}%
\bibitem [{\citenamefont {Gillespie}(2007)}]{gillespie2007stochastic}%
  \BibitemOpen
  \bibfield  {author} {\bibinfo {author} {\bibfnamefont {D.~T.}\ \bibnamefont
  {Gillespie}},\ }\href@noop {} {\bibfield  {journal} {\bibinfo  {journal}
  {Annu. Rev. Phys. Chem.}\ }\textbf {\bibinfo {volume} {58}},\ \bibinfo
  {pages} {35} (\bibinfo {year} {2007})}\BibitemShut {NoStop}%
\bibitem [{\citenamefont {Cao}, \citenamefont {Gillespie},\ and\ \citenamefont
  {Petzold}(2006)}]{cao2006efficient}%
  \BibitemOpen
  \bibfield  {author} {\bibinfo {author} {\bibfnamefont {Y.}~\bibnamefont
  {Cao}}, \bibinfo {author} {\bibfnamefont {D.~T.}\ \bibnamefont {Gillespie}},
  \ and\ \bibinfo {author} {\bibfnamefont {L.~R.}\ \bibnamefont {Petzold}},\
  }\href@noop {} {\bibfield  {journal} {\bibinfo  {journal} {The Journal of
  Chemical Physics}\ }\textbf {\bibinfo {volume} {124}},\ \bibinfo {pages}
  {044109} (\bibinfo {year} {2006})}\BibitemShut {NoStop}%
\bibitem [{\citenamefont {Fu}, \citenamefont {Wu},\ and\ \citenamefont
  {Petzold}(2013)}]{fu2013time}%
  \BibitemOpen
  \bibfield  {author} {\bibinfo {author} {\bibfnamefont {J.}~\bibnamefont
  {Fu}}, \bibinfo {author} {\bibfnamefont {S.}~\bibnamefont {Wu}}, \ and\
  \bibinfo {author} {\bibfnamefont {L.~R.}\ \bibnamefont {Petzold}},\
  }\href@noop {} {\bibfield  {journal} {\bibinfo  {journal} {Journal of
  Computational Physics}\ }\textbf {\bibinfo {volume} {235}},\ \bibinfo {pages}
  {446} (\bibinfo {year} {2013})}\BibitemShut {NoStop}%
\bibitem [{\citenamefont {Wu}, \citenamefont {Fu},\ and\ \citenamefont
  {Petzold}(2015)}]{wu2015adaptive}%
  \BibitemOpen
  \bibfield  {author} {\bibinfo {author} {\bibfnamefont {S.}~\bibnamefont
  {Wu}}, \bibinfo {author} {\bibfnamefont {J.}~\bibnamefont {Fu}}, \ and\
  \bibinfo {author} {\bibfnamefont {L.~R.}\ \bibnamefont {Petzold}},\
  }\href@noop {} {\bibfield  {journal} {\bibinfo  {journal} {The Journal of
  Chemical Physics}\ }\textbf {\bibinfo {volume} {142}},\ \bibinfo {pages}
  {204108} (\bibinfo {year} {2015})}\BibitemShut {NoStop}%
\bibitem [{\citenamefont {Flekk{\o}y}, \citenamefont {Feder},\ and\
  \citenamefont {Wagner}(2001)}]{flekkoy2001coupling}%
  \BibitemOpen
  \bibfield  {author} {\bibinfo {author} {\bibfnamefont {E.}~\bibnamefont
  {Flekk{\o}y}}, \bibinfo {author} {\bibfnamefont {J.}~\bibnamefont {Feder}}, \
  and\ \bibinfo {author} {\bibfnamefont {G.}~\bibnamefont {Wagner}},\
  }\href@noop {} {\bibfield  {journal} {\bibinfo  {journal} {Physical Review
  E}\ }\textbf {\bibinfo {volume} {64}},\ \bibinfo {pages} {066302} (\bibinfo
  {year} {2001})}\BibitemShut {NoStop}%
\bibitem [{\citenamefont {Alexander}, \citenamefont {Garcia},\ and\
  \citenamefont {Tartakovsky}(2002)}]{alexander2002algorithm}%
  \BibitemOpen
  \bibfield  {author} {\bibinfo {author} {\bibfnamefont {F.~J.}\ \bibnamefont
  {Alexander}}, \bibinfo {author} {\bibfnamefont {A.~L.}\ \bibnamefont
  {Garcia}}, \ and\ \bibinfo {author} {\bibfnamefont {D.~M.}\ \bibnamefont
  {Tartakovsky}},\ }\href@noop {} {\bibfield  {journal} {\bibinfo  {journal}
  {Journal of Computational Physics}\ }\textbf {\bibinfo {volume} {182}},\
  \bibinfo {pages} {47} (\bibinfo {year} {2002})}\BibitemShut {NoStop}%
\bibitem [{\citenamefont {Alexander}, \citenamefont {Garcia},\ and\
  \citenamefont {Tartakovsky}(2005)}]{alexander2005algorithm}%
  \BibitemOpen
  \bibfield  {author} {\bibinfo {author} {\bibfnamefont {F.~J.}\ \bibnamefont
  {Alexander}}, \bibinfo {author} {\bibfnamefont {A.~L.}\ \bibnamefont
  {Garcia}}, \ and\ \bibinfo {author} {\bibfnamefont {D.~M.}\ \bibnamefont
  {Tartakovsky}},\ }\href@noop {} {\bibfield  {journal} {\bibinfo  {journal}
  {Journal of Computational Physics}\ }\textbf {\bibinfo {volume} {207}},\
  \bibinfo {pages} {769} (\bibinfo {year} {2005})}\BibitemShut {NoStop}%
\bibitem [{\citenamefont {Li}, \citenamefont {Ebert},\ and\ \citenamefont
  {Hundsdorfer}(2012)}]{li2012spatially}%
  \BibitemOpen
  \bibfield  {author} {\bibinfo {author} {\bibfnamefont {C.}~\bibnamefont
  {Li}}, \bibinfo {author} {\bibfnamefont {U.}~\bibnamefont {Ebert}}, \ and\
  \bibinfo {author} {\bibfnamefont {W.}~\bibnamefont {Hundsdorfer}},\
  }\href@noop {} {\bibfield  {journal} {\bibinfo  {journal} {Journal of
  Computational Physics}\ }\textbf {\bibinfo {volume} {231}},\ \bibinfo {pages}
  {1020} (\bibinfo {year} {2012})}\BibitemShut {NoStop}%
\bibitem [{\citenamefont {Flegg}, \citenamefont {Chapman},\ and\ \citenamefont
  {Erban}(2011)}]{flegg2012}%
  \BibitemOpen
  \bibfield  {author} {\bibinfo {author} {\bibfnamefont {M.~B.}\ \bibnamefont
  {Flegg}}, \bibinfo {author} {\bibfnamefont {S.~J.}\ \bibnamefont {Chapman}},
  \ and\ \bibinfo {author} {\bibfnamefont {R.}~\bibnamefont {Erban}},\ }\href
  {\doibase 10.1098/rsif.2011.0574} {\bibfield  {journal} {\bibinfo  {journal}
  {Journal of The Royal Society Interface}\ } (\bibinfo {year} {2011}),\
  10.1098/rsif.2011.0574}\BibitemShut {NoStop}%
\bibitem [{\citenamefont {Franz}\ \emph {et~al.}(2013)\citenamefont {Franz},
  \citenamefont {Flegg}, \citenamefont {Chapman},\ and\ \citenamefont
  {Erban}}]{franz2013}%
  \BibitemOpen
  \bibfield  {author} {\bibinfo {author} {\bibfnamefont {B.}~\bibnamefont
  {Franz}}, \bibinfo {author} {\bibfnamefont {M.~B.}\ \bibnamefont {Flegg}},
  \bibinfo {author} {\bibfnamefont {S.~J.}\ \bibnamefont {Chapman}}, \ and\
  \bibinfo {author} {\bibfnamefont {R.}~\bibnamefont {Erban}},\ }\href@noop {}
  {\bibfield  {journal} {\bibinfo  {journal} {SIAM J. Appl. Math.}\ }\textbf
  {\bibinfo {volume} {73}},\ \bibinfo {pages} {1224} (\bibinfo {year}
  {2013})}\BibitemShut {NoStop}%
\bibitem [{\citenamefont {Flegg}, \citenamefont {Hellander},\ and\
  \citenamefont {Erban}(2015)}]{flegg2015convergence}%
  \BibitemOpen
  \bibfield  {author} {\bibinfo {author} {\bibfnamefont {M.~B.}\ \bibnamefont
  {Flegg}}, \bibinfo {author} {\bibfnamefont {S.}~\bibnamefont {Hellander}}, \
  and\ \bibinfo {author} {\bibfnamefont {R.}~\bibnamefont {Erban}},\
  }\href@noop {} {\bibfield  {journal} {\bibinfo  {journal} {Journal of
  Computational Physics}\ }\textbf {\bibinfo {volume} {289}},\ \bibinfo {pages}
  {1} (\bibinfo {year} {2015})}\BibitemShut {NoStop}%
\bibitem [{\citenamefont {Robinson}, \citenamefont {Flegg},\ and\ \citenamefont
  {Erban}(2014)}]{robinson2014adaptive}%
  \BibitemOpen
  \bibfield  {author} {\bibinfo {author} {\bibfnamefont {M.}~\bibnamefont
  {Robinson}}, \bibinfo {author} {\bibfnamefont {M.}~\bibnamefont {Flegg}}, \
  and\ \bibinfo {author} {\bibfnamefont {R.}~\bibnamefont {Erban}},\
  }\href@noop {} {\bibfield  {journal} {\bibinfo  {journal} {The Journal of
  Chemical Physics}\ }\textbf {\bibinfo {volume} {140}},\ \bibinfo {pages}
  {124109} (\bibinfo {year} {2014})}\BibitemShut {NoStop}%
\bibitem [{\citenamefont {Sanft}\ and\ \citenamefont
  {Othmer}(2015)}]{sanft2015constant}%
  \BibitemOpen
  \bibfield  {author} {\bibinfo {author} {\bibfnamefont {K.~R.}\ \bibnamefont
  {Sanft}}\ and\ \bibinfo {author} {\bibfnamefont {H.~G.}\ \bibnamefont
  {Othmer}},\ }\href@noop {} {\bibfield  {journal} {\bibinfo  {journal} {arXiv
  preprint arXiv:1503.05832}\ } (\bibinfo {year} {2015})}\BibitemShut {NoStop}%
\bibitem [{\citenamefont {Yates}\ and\ \citenamefont
  {Flegg}(2015)}]{yates2015pseudo}%
  \BibitemOpen
  \bibfield  {author} {\bibinfo {author} {\bibfnamefont {C.~A.}\ \bibnamefont
  {Yates}}\ and\ \bibinfo {author} {\bibfnamefont {M.~B.}\ \bibnamefont
  {Flegg}},\ }\href {\doibase 10.1098/rsif.2015.0141} {\bibfield  {journal}
  {\bibinfo  {journal} {Journal of The Royal Society Interface}\ }\textbf
  {\bibinfo {volume} {12}} (\bibinfo {year} {2015}),\
  10.1098/rsif.2015.0141}\BibitemShut {NoStop}%
\bibitem [{\citenamefont {Bezzola}\ \emph {et~al.}(2014)\citenamefont
  {Bezzola}, \citenamefont {Bales}, \citenamefont {Alkire},\ and\ \citenamefont
  {Petzold}}]{bezzola2014exact}%
  \BibitemOpen
  \bibfield  {author} {\bibinfo {author} {\bibfnamefont {A.}~\bibnamefont
  {Bezzola}}, \bibinfo {author} {\bibfnamefont {B.~B.}\ \bibnamefont {Bales}},
  \bibinfo {author} {\bibfnamefont {R.~C.}\ \bibnamefont {Alkire}}, \ and\
  \bibinfo {author} {\bibfnamefont {L.~R.}\ \bibnamefont {Petzold}},\
  }\href@noop {} {\bibfield  {journal} {\bibinfo  {journal} {Journal of
  Computational Physics}\ }\textbf {\bibinfo {volume} {256}},\ \bibinfo {pages}
  {183} (\bibinfo {year} {2014})}\BibitemShut {NoStop}%
\bibitem [{\citenamefont {Breuer}, \citenamefont {Huber},\ and\ \citenamefont
  {Petruccione}(1994)}]{breuer1994fluctuation}%
  \BibitemOpen
  \bibfield  {author} {\bibinfo {author} {\bibfnamefont {H.-P.}\ \bibnamefont
  {Breuer}}, \bibinfo {author} {\bibfnamefont {W.}~\bibnamefont {Huber}}, \
  and\ \bibinfo {author} {\bibfnamefont {F.}~\bibnamefont {Petruccione}},\
  }\href@noop {} {\bibfield  {journal} {\bibinfo  {journal} {Physica D:
  Nonlinear Phenomena}\ }\textbf {\bibinfo {volume} {73}},\ \bibinfo {pages}
  {259} (\bibinfo {year} {1994})}\BibitemShut {NoStop}%
\bibitem [{\citenamefont {Huang}, \citenamefont {Meir},\ and\ \citenamefont
  {Wingreen}(2003)}]{huang2003dynamic}%
  \BibitemOpen
  \bibfield  {author} {\bibinfo {author} {\bibfnamefont {K.~C.}\ \bibnamefont
  {Huang}}, \bibinfo {author} {\bibfnamefont {Y.}~\bibnamefont {Meir}}, \ and\
  \bibinfo {author} {\bibfnamefont {N.~S.}\ \bibnamefont {Wingreen}},\
  }\href@noop {} {\bibfield  {journal} {\bibinfo  {journal} {Proceedings of the
  National Academy of Sciences}\ }\textbf {\bibinfo {volume} {100}},\ \bibinfo
  {pages} {12724} (\bibinfo {year} {2003})}\BibitemShut {NoStop}%
\bibitem [{\citenamefont {Kerr}\ \emph {et~al.}(2006)\citenamefont {Kerr},
  \citenamefont {Levine}, \citenamefont {Sejnowski},\ and\ \citenamefont
  {Rappel}}]{kerr2006division}%
  \BibitemOpen
  \bibfield  {author} {\bibinfo {author} {\bibfnamefont {R.~A.}\ \bibnamefont
  {Kerr}}, \bibinfo {author} {\bibfnamefont {H.}~\bibnamefont {Levine}},
  \bibinfo {author} {\bibfnamefont {T.~J.}\ \bibnamefont {Sejnowski}}, \ and\
  \bibinfo {author} {\bibfnamefont {W.-J.}\ \bibnamefont {Rappel}},\
  }\href@noop {} {\bibfield  {journal} {\bibinfo  {journal} {Proceedings of the
  National Academy of Sciences of the United States of America}\ }\textbf
  {\bibinfo {volume} {103}},\ \bibinfo {pages} {347} (\bibinfo {year}
  {2006})}\BibitemShut {NoStop}%
\bibitem [{\citenamefont {Fange}\ and\ \citenamefont
  {Elf}(2006)}]{fange2006noise}%
  \BibitemOpen
  \bibfield  {author} {\bibinfo {author} {\bibfnamefont {D.}~\bibnamefont
  {Fange}}\ and\ \bibinfo {author} {\bibfnamefont {J.}~\bibnamefont {Elf}},\
  }\href@noop {} {\bibfield  {journal} {\bibinfo  {journal} {PLoS Comput Biol}\
  }\textbf {\bibinfo {volume} {2}},\ \bibinfo {pages} {e80} (\bibinfo {year}
  {2006})}\BibitemShut {NoStop}%
\bibitem [{\citenamefont {Isaacson}(2013)}]{isaacson2013convergent}%
  \BibitemOpen
  \bibfield  {author} {\bibinfo {author} {\bibfnamefont {S.~A.}\ \bibnamefont
  {Isaacson}},\ }\href@noop {} {\bibfield  {journal} {\bibinfo  {journal} {J.
  Chem. Phys.}\ }\textbf {\bibinfo {volume} {139}},\ \bibinfo {pages} {054101}
  (\bibinfo {year} {2013})}\BibitemShut {NoStop}%
\bibitem [{\citenamefont {Gillespie}, \citenamefont {Petzold},\ and\
  \citenamefont {Seitaridou}(2014)}]{gillespie2014validity}%
  \BibitemOpen
  \bibfield  {author} {\bibinfo {author} {\bibfnamefont {D.~T.}\ \bibnamefont
  {Gillespie}}, \bibinfo {author} {\bibfnamefont {L.~R.}\ \bibnamefont
  {Petzold}}, \ and\ \bibinfo {author} {\bibfnamefont {E.}~\bibnamefont
  {Seitaridou}},\ }\href@noop {} {\bibfield  {journal} {\bibinfo  {journal}
  {The Journal of Chemical Physics}\ }\textbf {\bibinfo {volume} {140}},\
  \bibinfo {pages} {054111} (\bibinfo {year} {2014})}\BibitemShut {NoStop}%
\bibitem [{\citenamefont {Hellander}, \citenamefont {Hellander},\ and\
  \citenamefont {Petzold}(2015)}]{hellander2015reaction}%
  \BibitemOpen
  \bibfield  {author} {\bibinfo {author} {\bibfnamefont {S.}~\bibnamefont
  {Hellander}}, \bibinfo {author} {\bibfnamefont {A.}~\bibnamefont
  {Hellander}}, \ and\ \bibinfo {author} {\bibfnamefont {L.}~\bibnamefont
  {Petzold}},\ }\href@noop {} {\bibfield  {journal} {\bibinfo  {journal}
  {Physical Review E}\ }\textbf {\bibinfo {volume} {91}},\ \bibinfo {pages}
  {023312} (\bibinfo {year} {2015})}\BibitemShut {NoStop}%
\bibitem [{\citenamefont {Isaacson}\ and\ \citenamefont
  {Peskin}(2006)}]{isaacson2006incorporating}%
  \BibitemOpen
  \bibfield  {author} {\bibinfo {author} {\bibfnamefont {S.~A.}\ \bibnamefont
  {Isaacson}}\ and\ \bibinfo {author} {\bibfnamefont {C.~S.}\ \bibnamefont
  {Peskin}},\ }\href@noop {} {\bibfield  {journal} {\bibinfo  {journal} {SIAM
  Journal on Scientific Computing}\ }\textbf {\bibinfo {volume} {28}},\
  \bibinfo {pages} {47} (\bibinfo {year} {2006})}\BibitemShut {NoStop}%
\bibitem [{\citenamefont {Drawert}, \citenamefont {Engblom},\ and\
  \citenamefont {Hellander}(2012)}]{drawert2012urdme}%
  \BibitemOpen
  \bibfield  {author} {\bibinfo {author} {\bibfnamefont {B.}~\bibnamefont
  {Drawert}}, \bibinfo {author} {\bibfnamefont {S.}~\bibnamefont {Engblom}}, \
  and\ \bibinfo {author} {\bibfnamefont {A.}~\bibnamefont {Hellander}},\
  }\href@noop {} {\bibfield  {journal} {\bibinfo  {journal} {BMC Systems
  Biology}\ }\textbf {\bibinfo {volume} {6}},\ \bibinfo {pages} {76} (\bibinfo
  {year} {2012})}\BibitemShut {NoStop}%
\end{thebibliography}

\pagebreak

\beginsupplement

\section{Supplementary Information}

\begin{figure}[h!]

\subfloat[]
{\label{fig:BirthDeathDistribution1}
\includegraphics[width=0.48\linewidth]{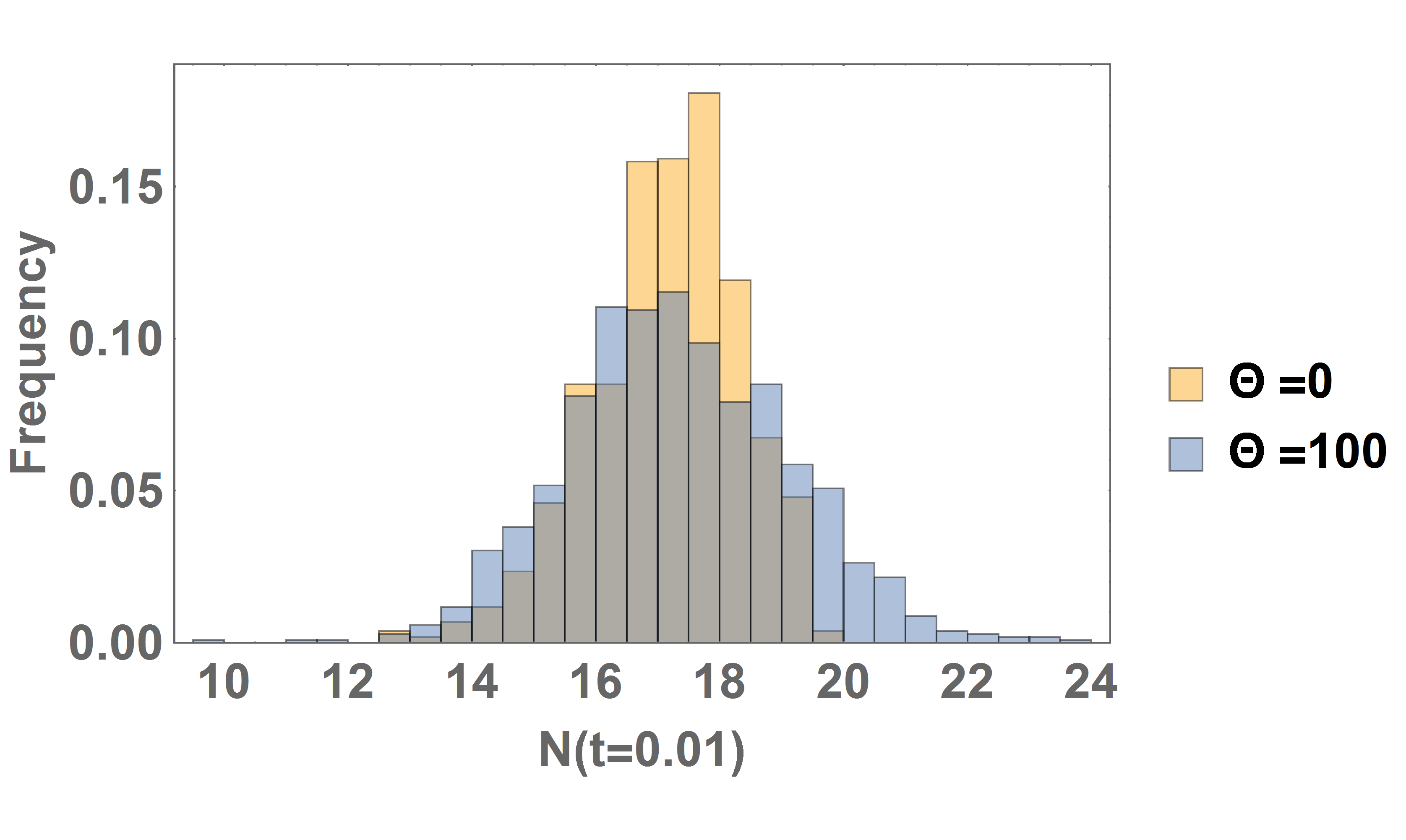}
}
\subfloat[]
{\label{fig:BirthDeathDistribution2}
\includegraphics[width=0.48\linewidth]{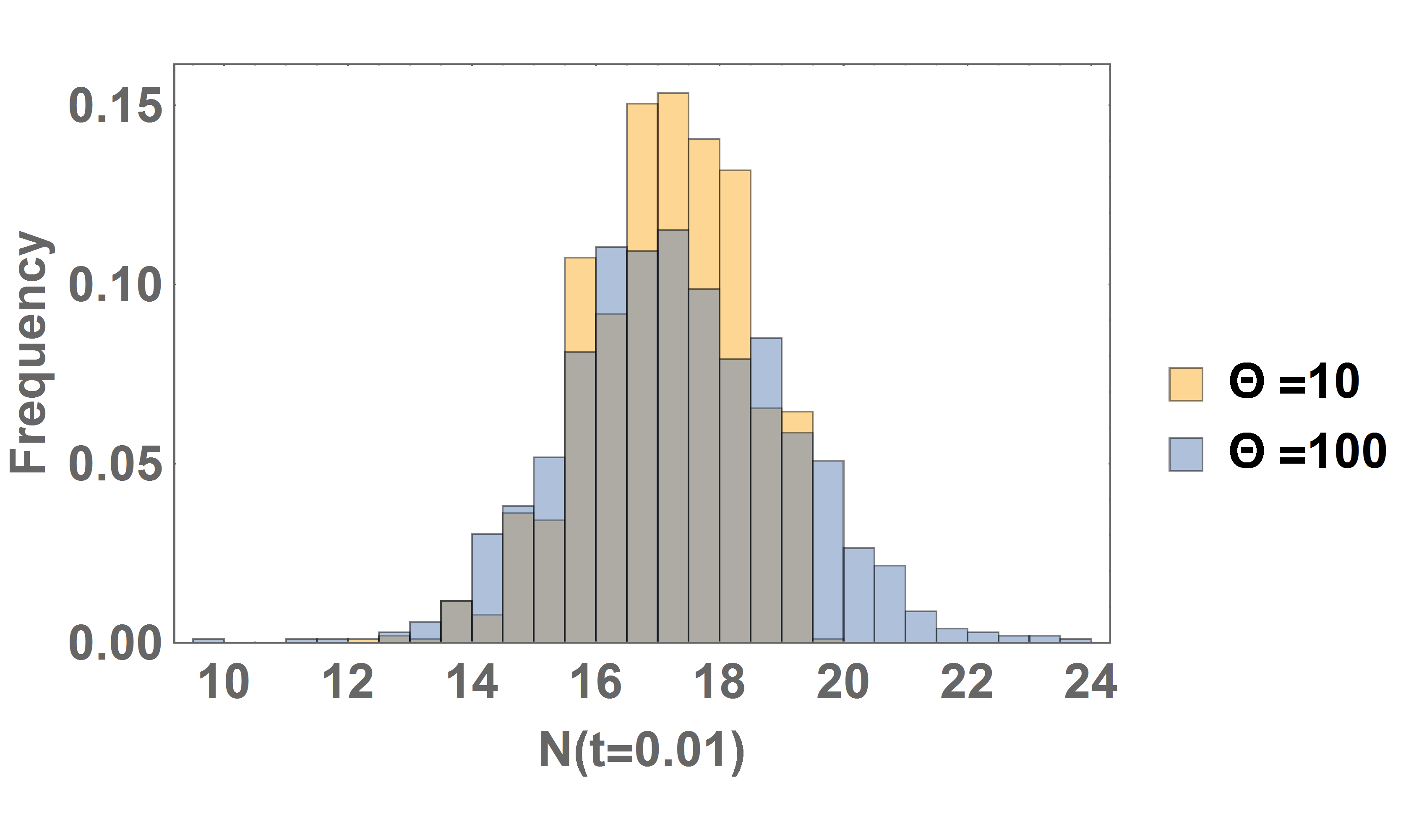}
}\\
\subfloat[]
{\label{fig:BirthDeathDistribution3}
\includegraphics[width=0.48\linewidth]{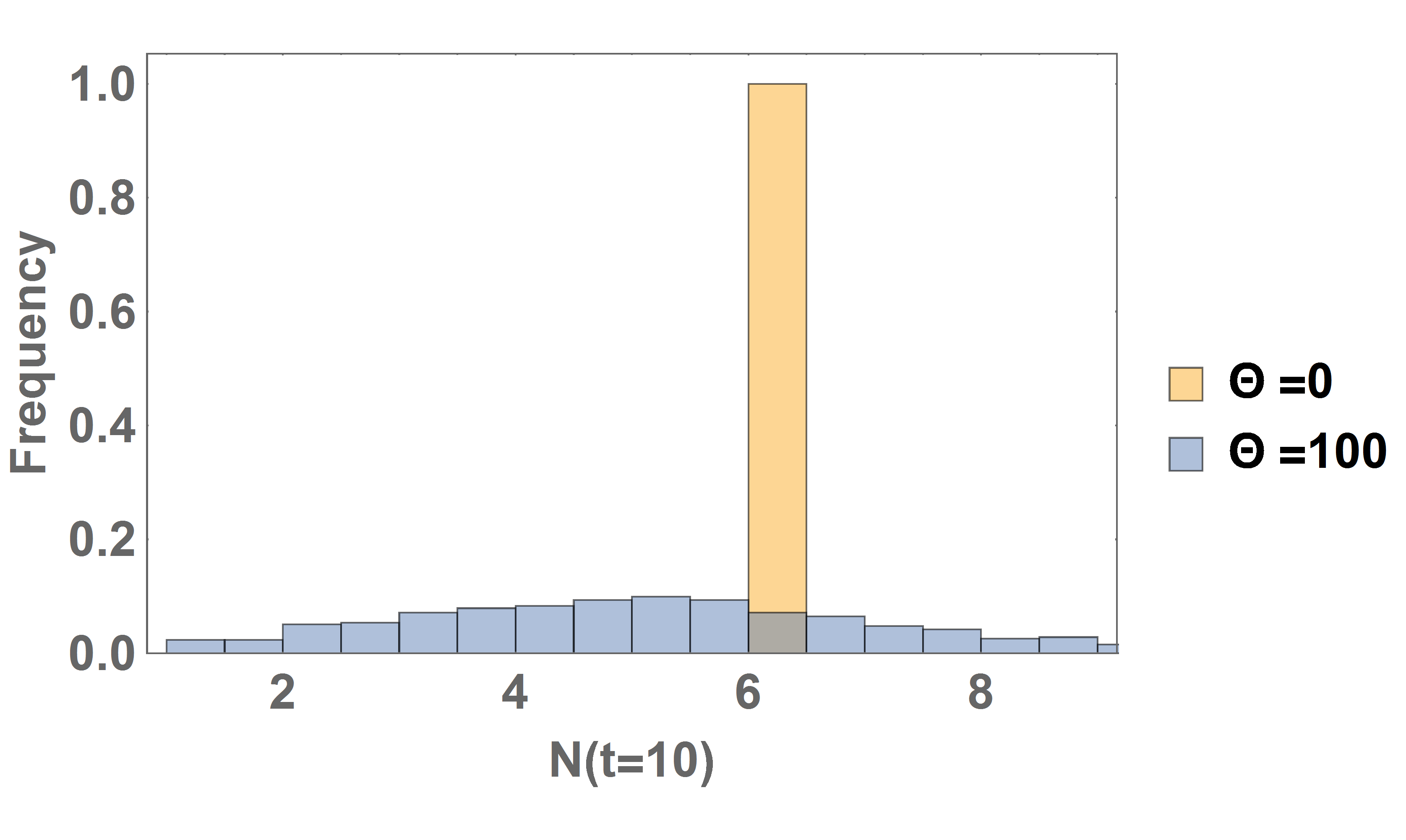}
}
\subfloat[]
{\label{fig:BirthDeathDistribution4}
\includegraphics[width=0.48\linewidth]{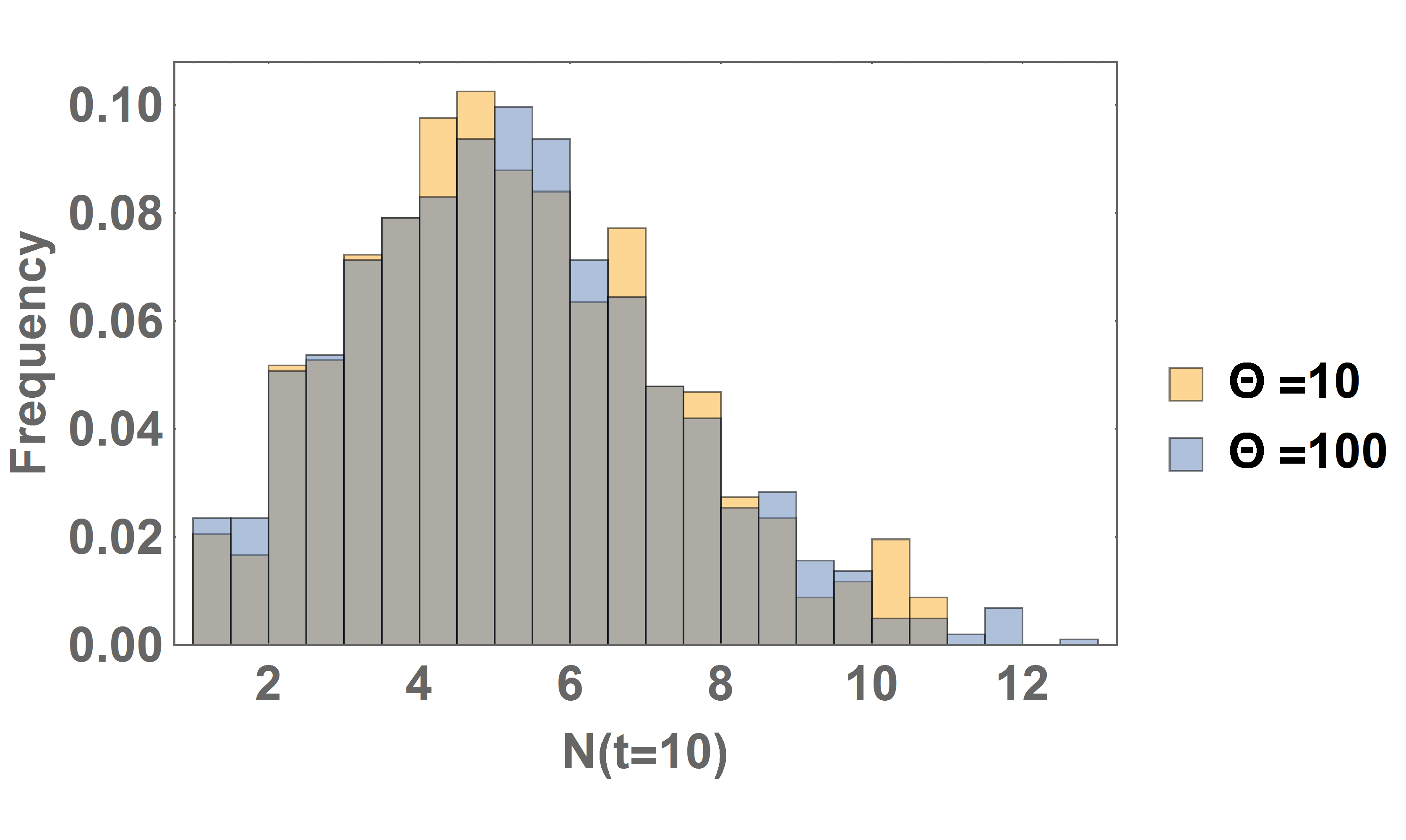}
}
\caption{Distribution of the birth-death process for the same parameters as in Figure \ref{fig:BirthDeath}, obtained from running $1024$ different simulations, for two thresholds $\Theta = 0, 10$. In each case, we compare to the exact model, obtained by choosing a threshold $\Theta=100$. The initial condition was $N(t=0) = 20$, and the steady state is at $N=6$. We notice that at $t=0.01$, the distributions of all three cases, $\Theta = 0,10,100$ are similar, but, by construction, for $\Theta<20$, no birth can occur so the distribution is bounded by $N(t=0)$. At $t=10$, the distributions for $\Theta=10, 100$ look indistinguishable, whereas the distribution for $\Theta=0$ goes to the delta distribution, and the solution is trapped at the absorbing state.
\label{fig:BirthDeathDistribution}} 
\end{figure}

\pagebreak

\begin{figure}[h!]

\subfloat[]
{\label{fig:DiffusionDistribution1}
\includegraphics[width=0.48\linewidth]{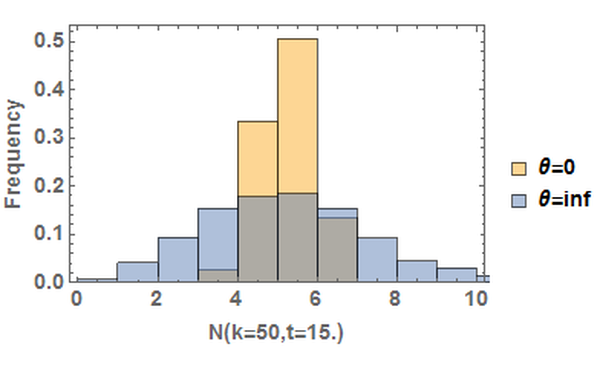}
}
\subfloat[]
{\label{fig:DiffusionDistribution2}
\includegraphics[width=0.48\linewidth]{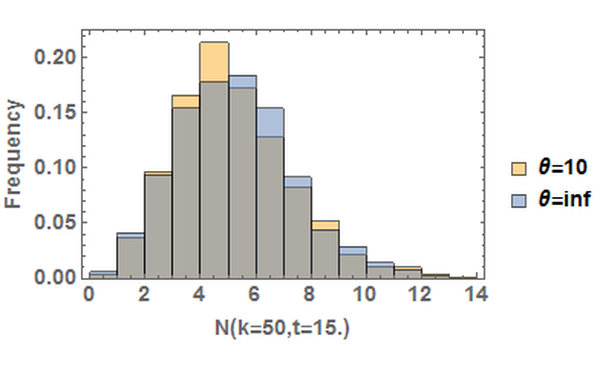}
}\\
\subfloat[]
{\label{fig:DiffusionDistribution3}
\includegraphics[width=0.48\linewidth]{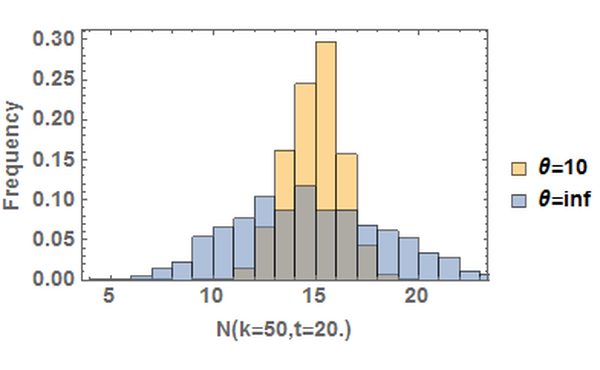}
}
\subfloat[]
{\label{fig:DiffusionDistribution4}
\includegraphics[width=0.48\linewidth]{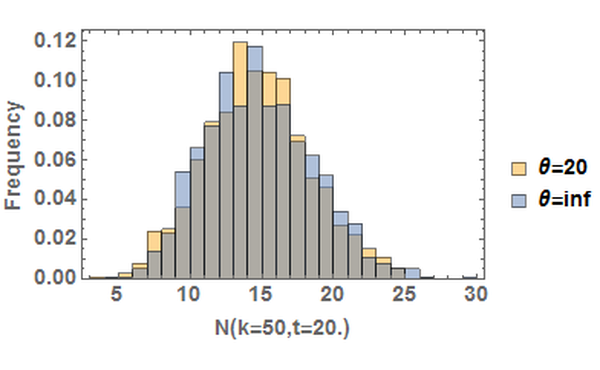}
}
\caption{Distribution of number of particles in compartment $k=50$ for the diffusion process described in section \ref{sec:firstPassageTime}, where initially $1000$ particles are seeded in the compartment on the left, with all other compartments empty. After $t=15$, we see that, while the conditional difference model with a threshold of $\Theta=10$ closely matches the exact model ($\Theta=inf$), panel \protect\subref{fig:DiffusionDistribution2}, the model with a threshold of $\Theta=0$ shows a much narrower distribution than the exact model (panel \protect\subref{fig:DiffusionDistribution1}). At time $t=20$, the case of $\Theta=10$ now also deviates visibly from the exact case (panel \protect\subref{fig:DiffusionDistribution3}), whereas the case of  $\Theta=20$ offers close agreement (panel \protect\subref{fig:DiffusionDistribution4}).
\label{fig:DiffusionDistribution}} 
\end{figure}

\pagebreak

\begin{figure}[h!]
\centering
\includegraphics[width=0.8\linewidth]{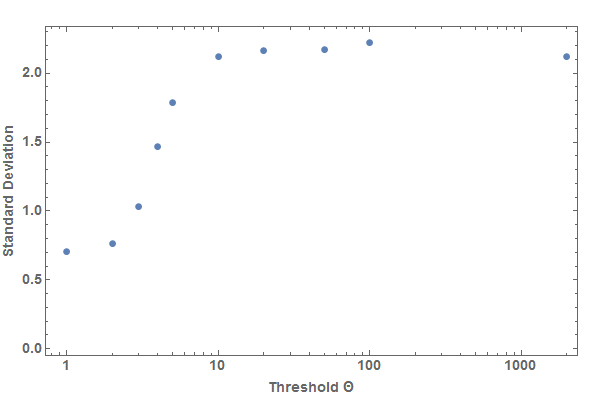}
\caption{
Standard Deviation of the diffusion process described in section \ref{sec:firstPassageTime}, obtained from distributions similar to to those shown in Figure \ref{fig:DiffusionDistribution}. Here, measurements are taken at fixed time $t=5$ in compartment $k=30$, from $1024$ simulations. We see that the standard deviations approach the ones of the exact model for thresholds around $\Theta =10$. 
\label{fig:DiffusionDistributionMoments}
} 
\end{figure}

\begin{figure}[h!]
\includegraphics[width=0.6\linewidth]{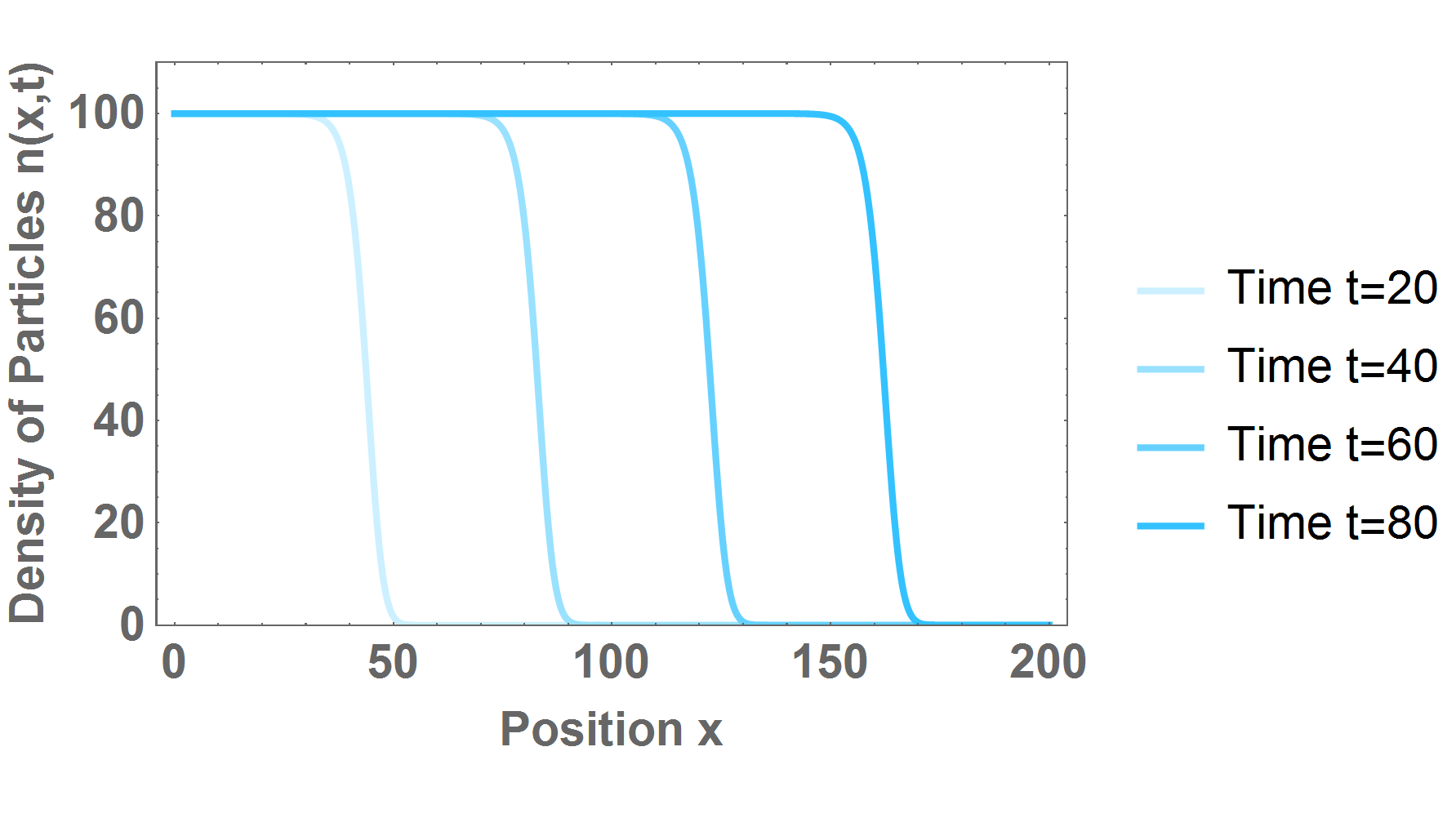}
\caption{
Travelling waves of the deterministic Fisher-Kolmogorov equation for parameters $D=1$, $\lambda=1$ and $\Omega=100$. We see that a traveling wave is spreading with constant velocity $v=2\sqrt{D\lambda} = 2$.
\label{fig:FisherDeterministic}
} 
\end{figure}

\pagebreak

\pagebreak
\begin{figure}[h!]

\subfloat[]
{\label{fig:MinMeanMinDADP}
\includegraphics[width=0.48\linewidth]{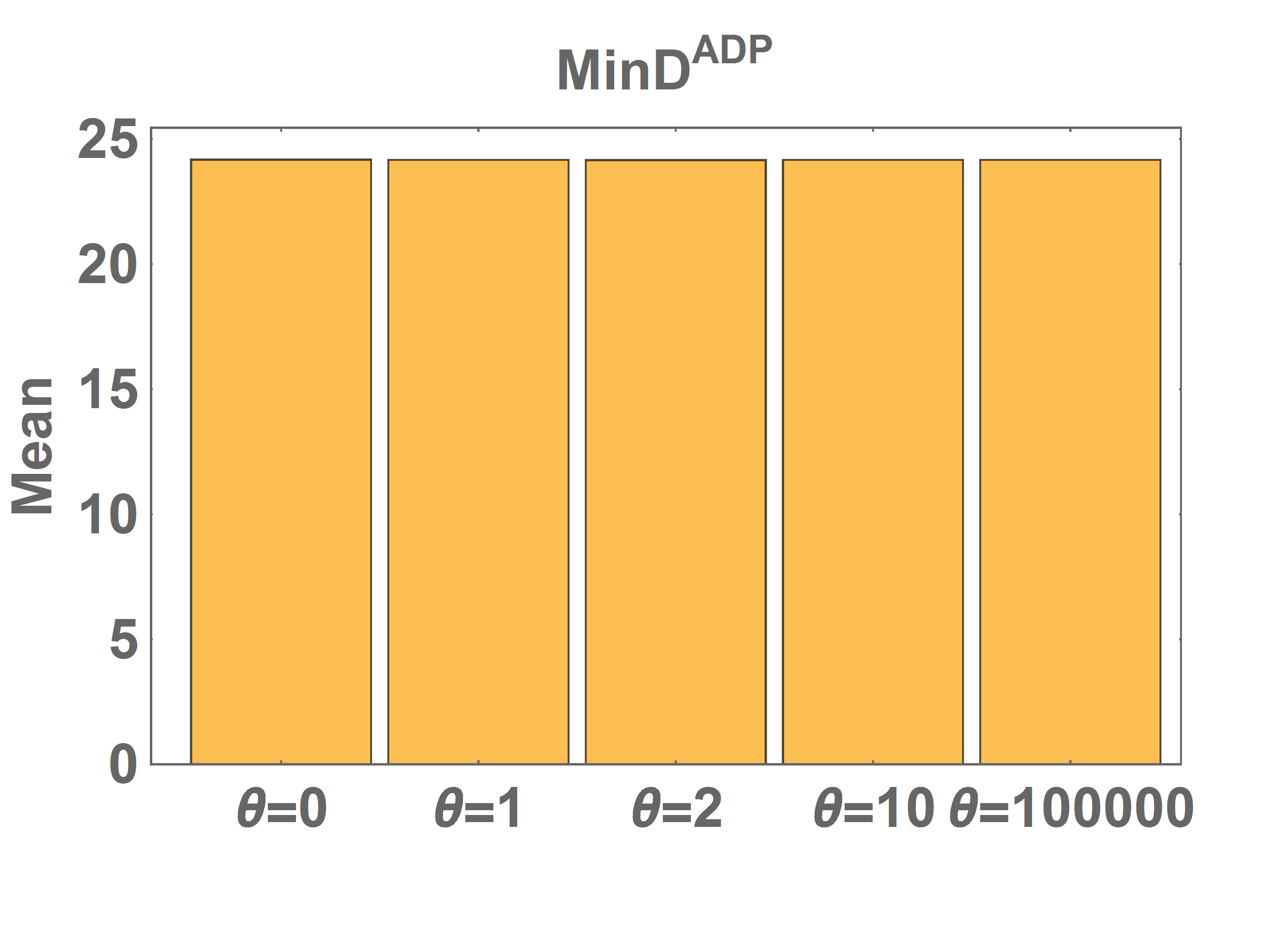}
}
\subfloat[]
{\label{fig:MinMeanMinDATP}
\includegraphics[width=0.48\linewidth]{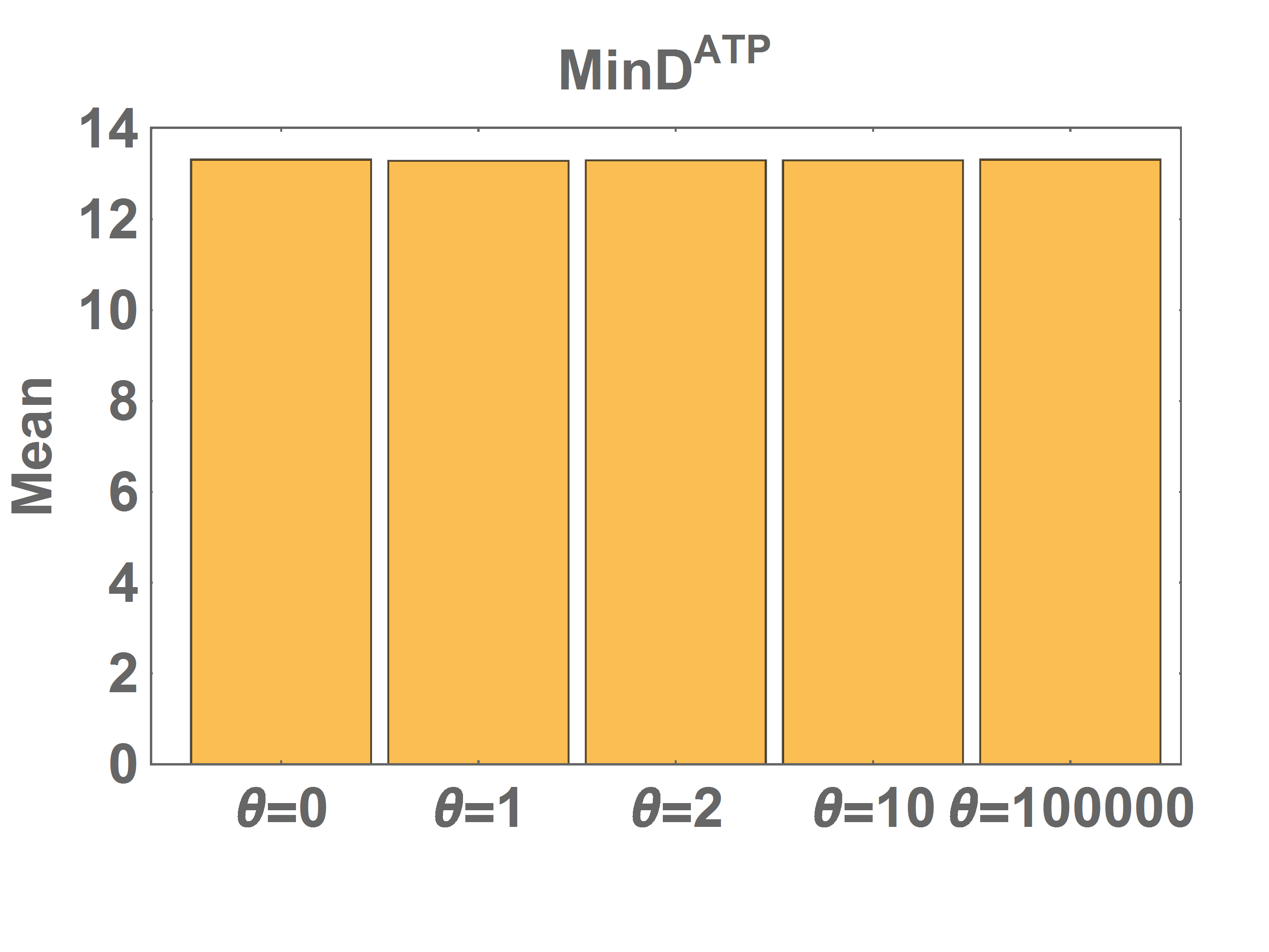}
}\\
\subfloat[]
{\label{fig:MinMeanMinE}
\includegraphics[width=0.48\linewidth]{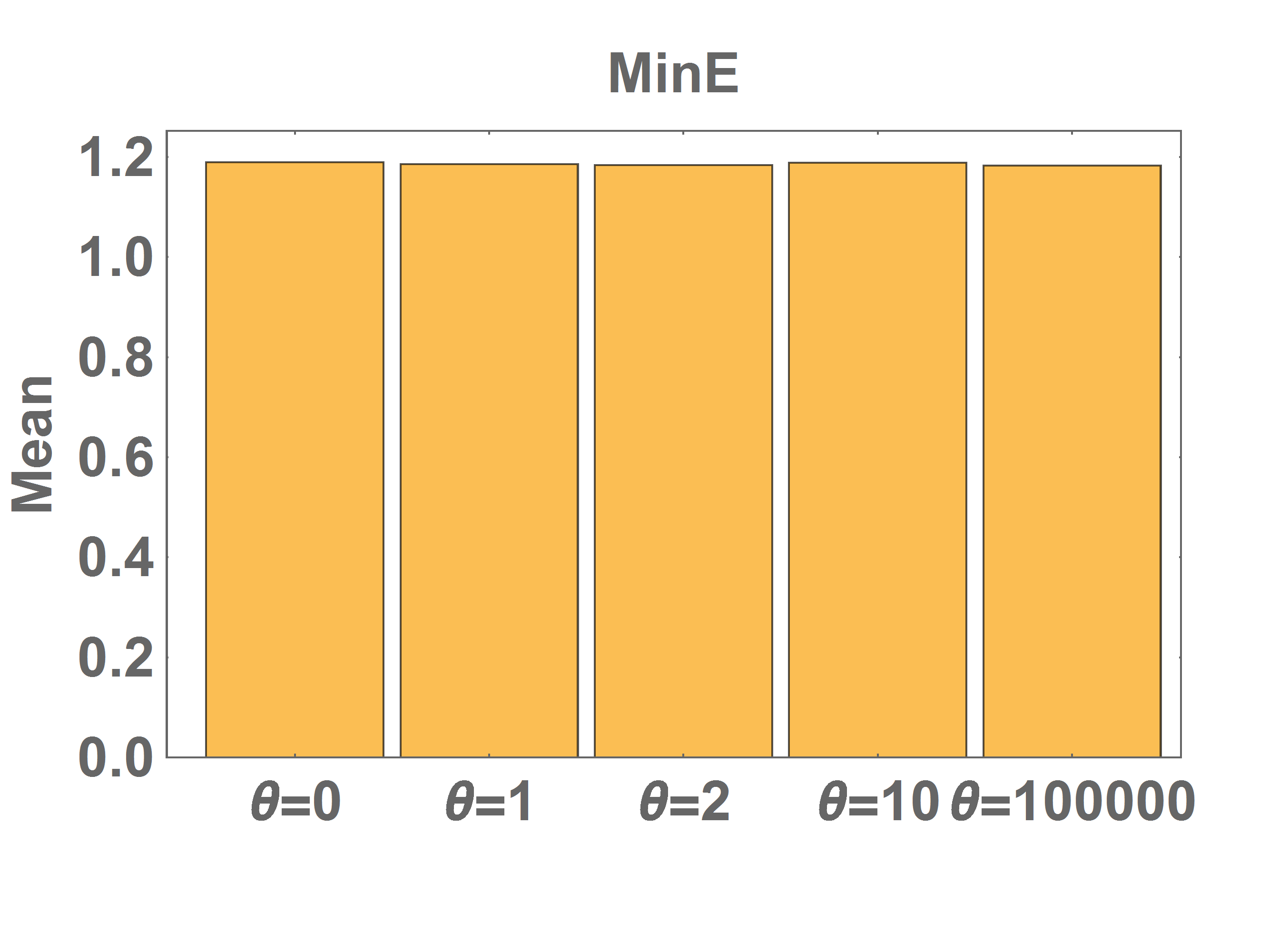}
}
\subfloat[]
{\label{fig:MinMeanMinDM}
\includegraphics[width=0.48\linewidth]{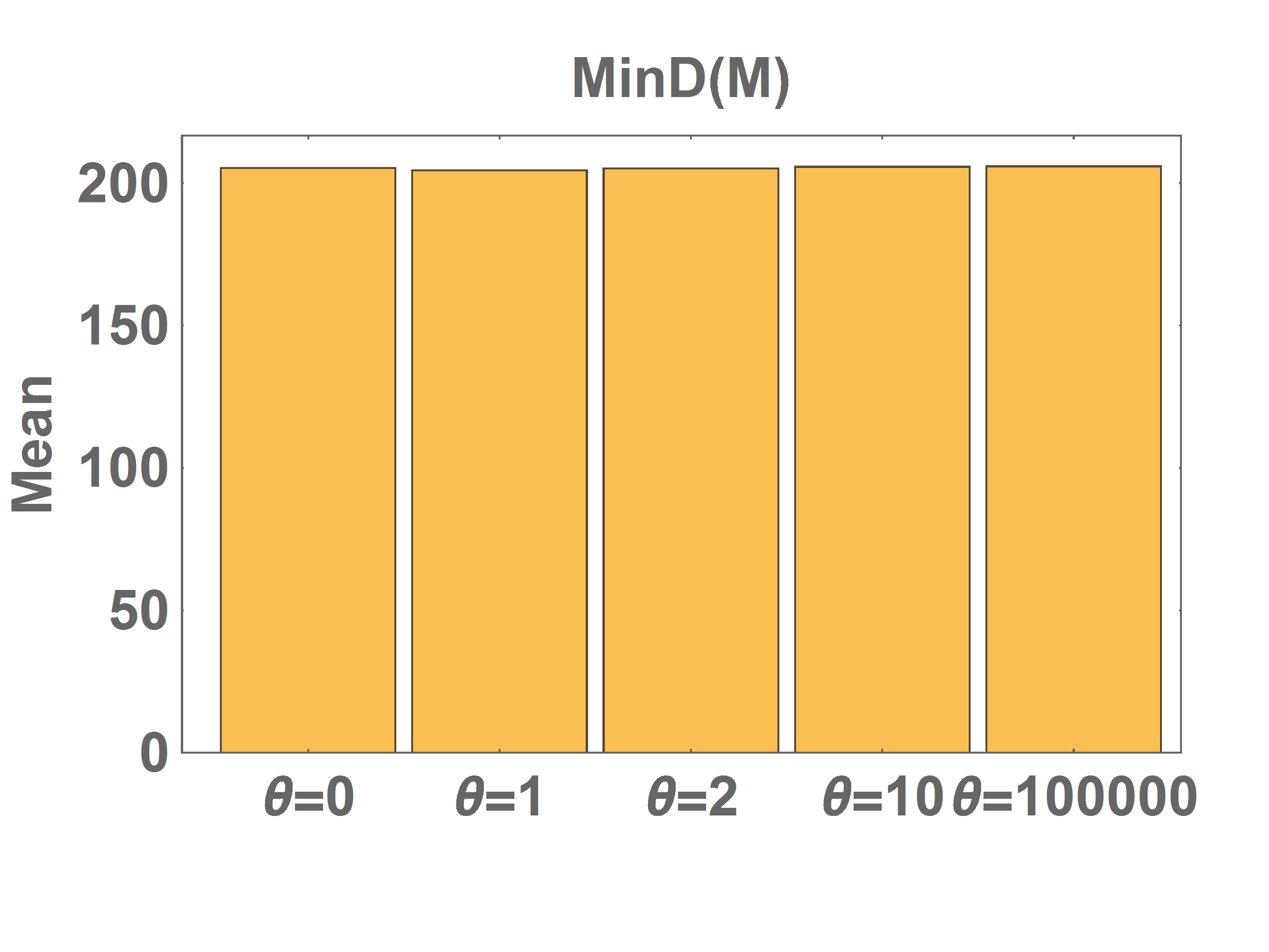}
}
\caption{Mean of the distributions shown in \ref{fig:Histogram}, corresponding to the standard deviations shown in Figure \ref{fig:MinStandardDeviations}.
\label{fig:MinMean}} 
\end{figure}
\pagebreak
\begin{figure}[h!]

\subfloat[]
{\label{fig:MinStandardDeviationsMinDAD_350P}
\includegraphics[width=0.48\linewidth]{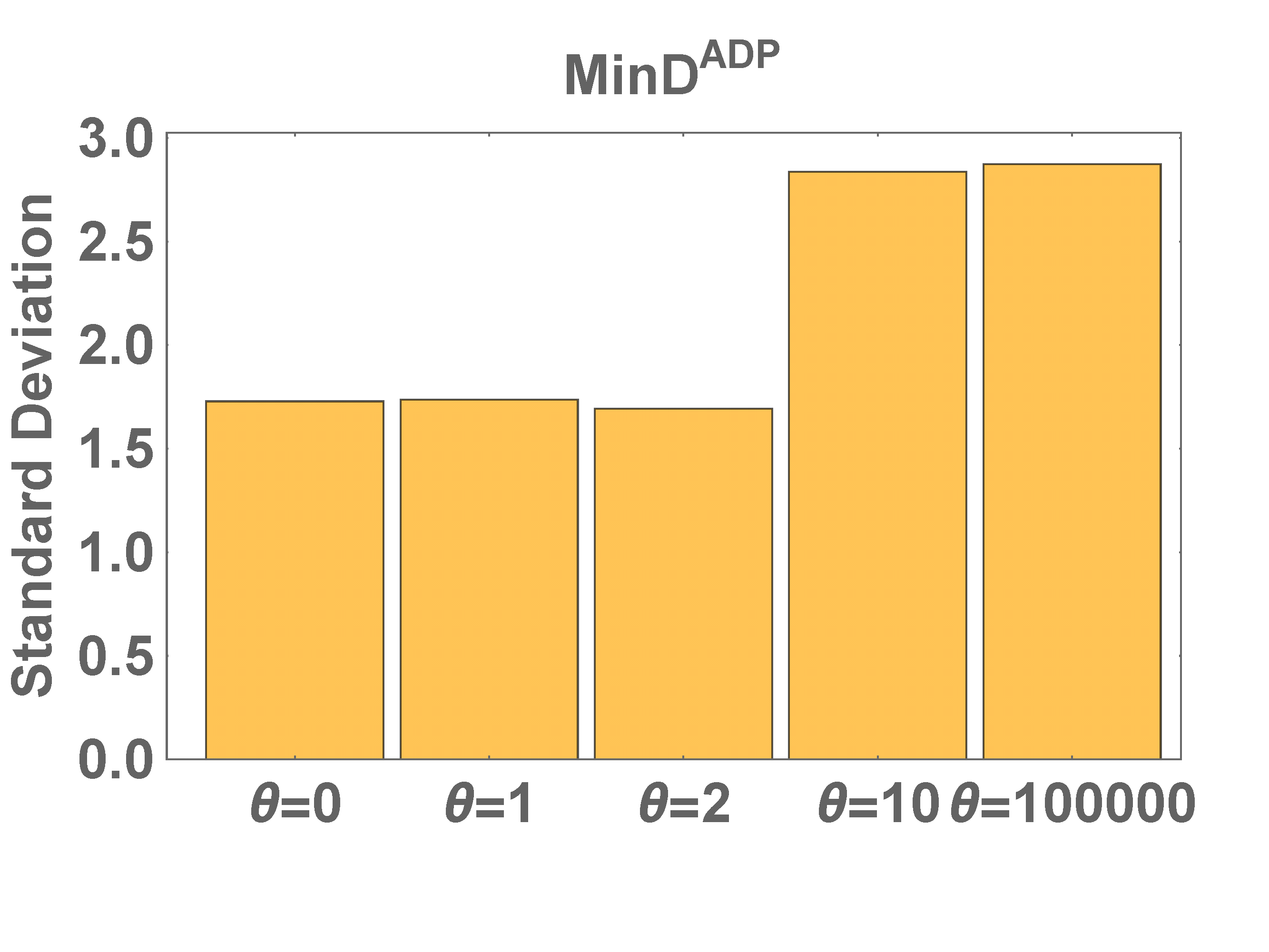}
}
\subfloat[]
{\label{fig:MinStandardDeviationsMinDATP_350}
\includegraphics[width=0.48\linewidth]{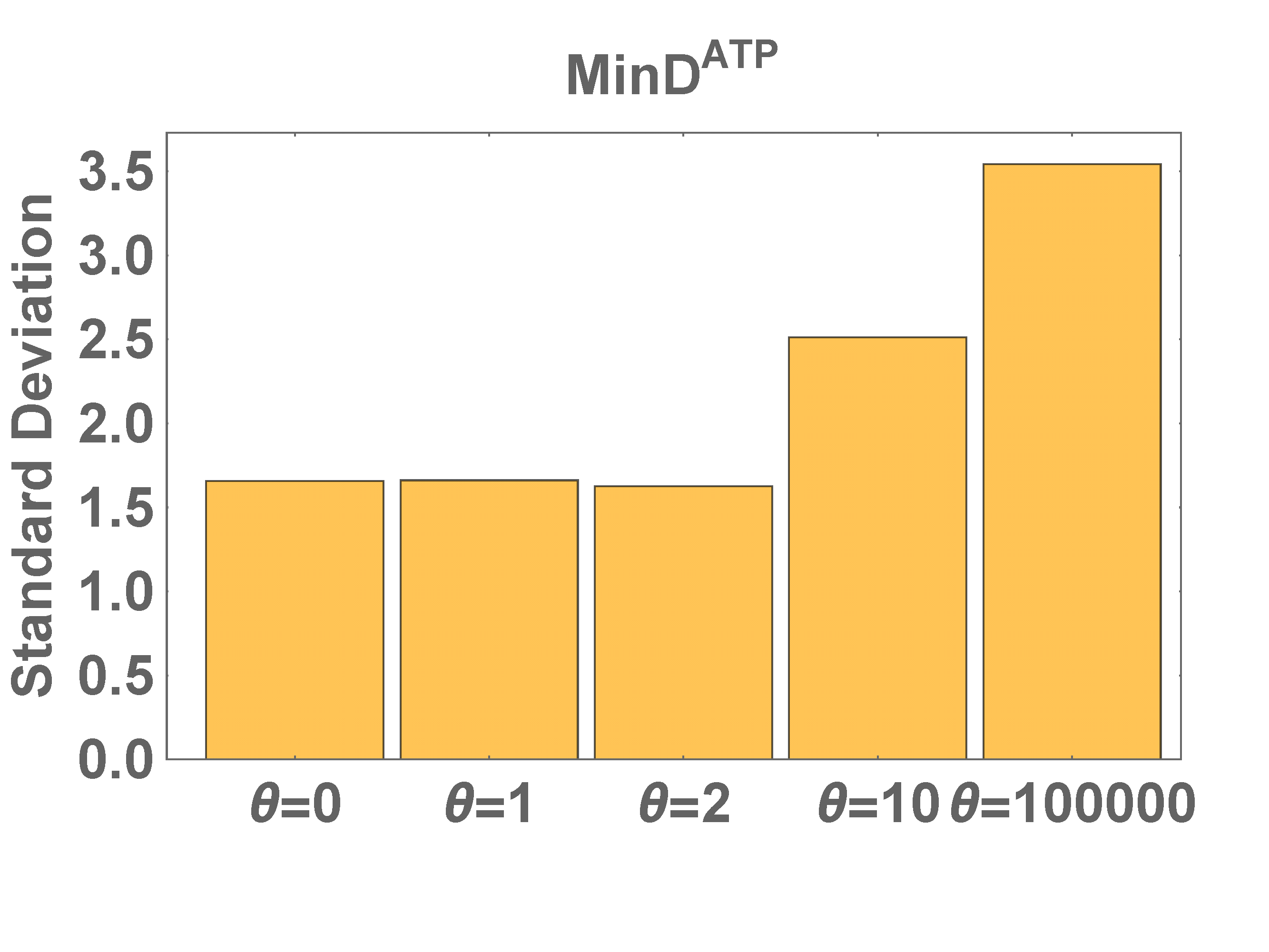}
}\\
\subfloat[]
{\label{fig:MinStandardDeviationsMinE_350}
\includegraphics[width=0.48\linewidth]{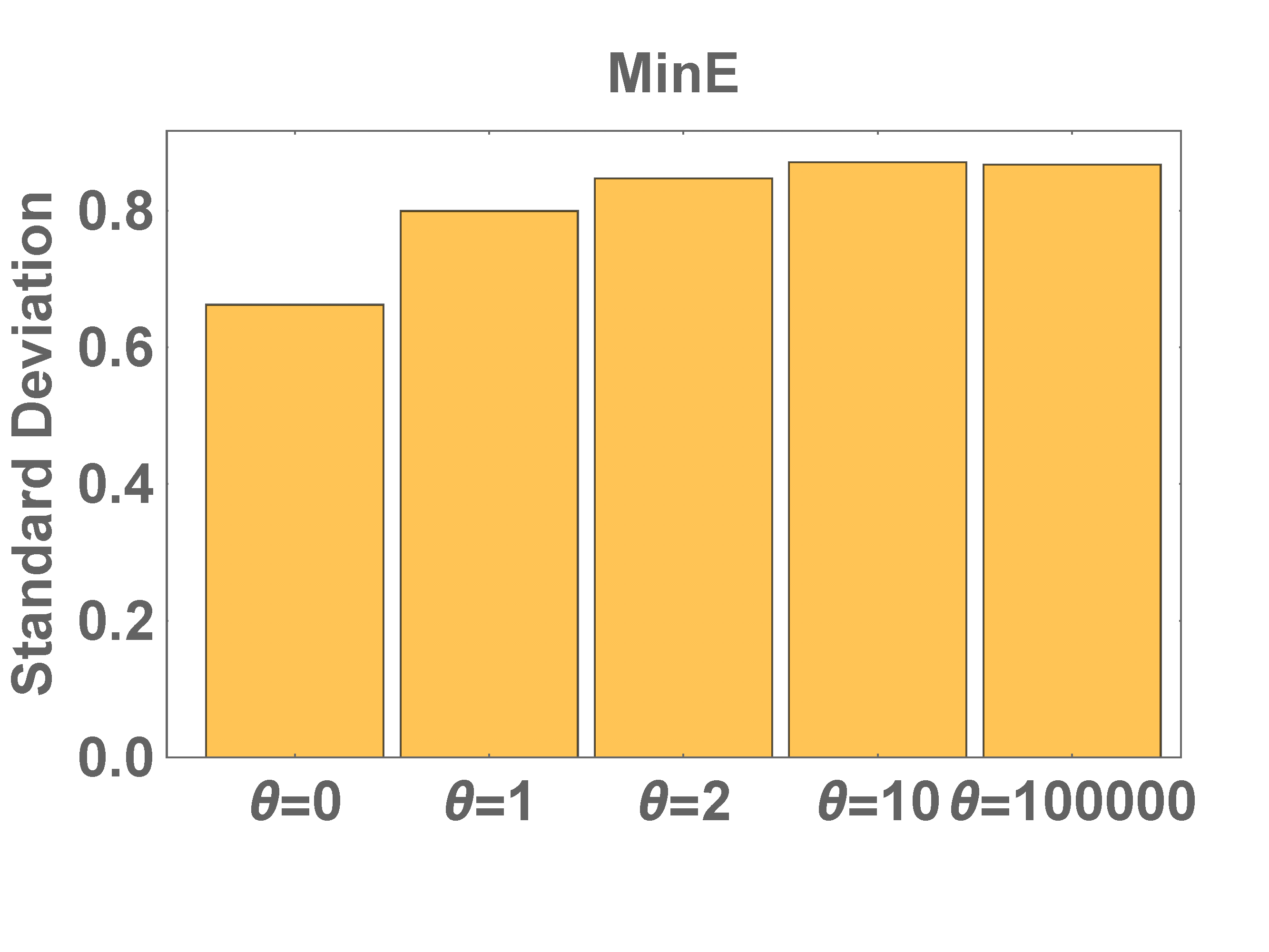}
}
\subfloat[]
{\label{fig:MinStandardDeviationsMinDM_350}
\includegraphics[width=0.48\linewidth]{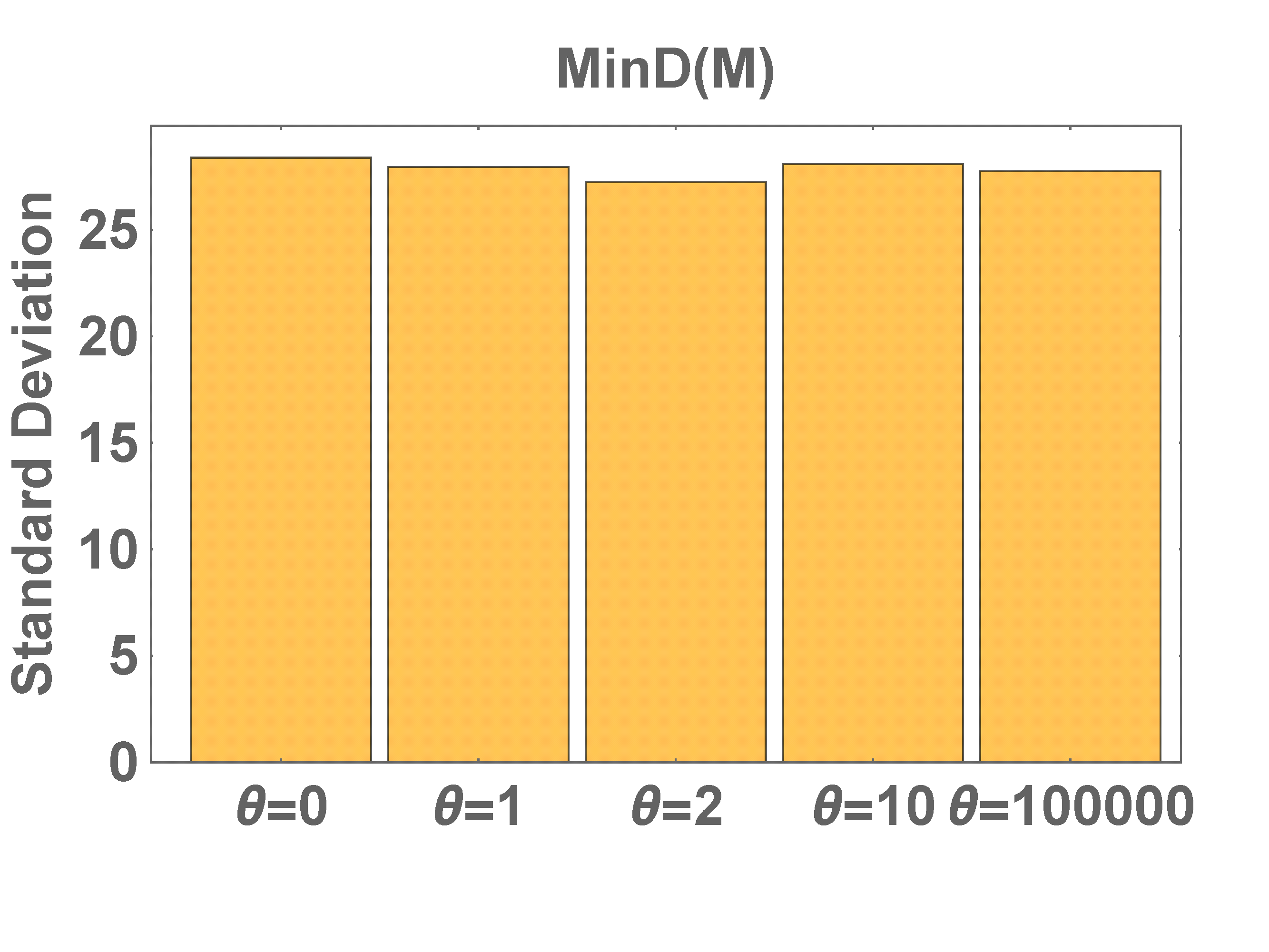}
}
\caption{Standard deviations from the distributions of particles, as in Figure \ref{fig:MinStandardDeviations}, but for $\frac{1}{4}$ of the number of particles.
\label{fig:MinStandardDeviations_350}} 
\end{figure}

\pagebreak
\begin{figure}[h!]
\includegraphics[width=0.48\linewidth]{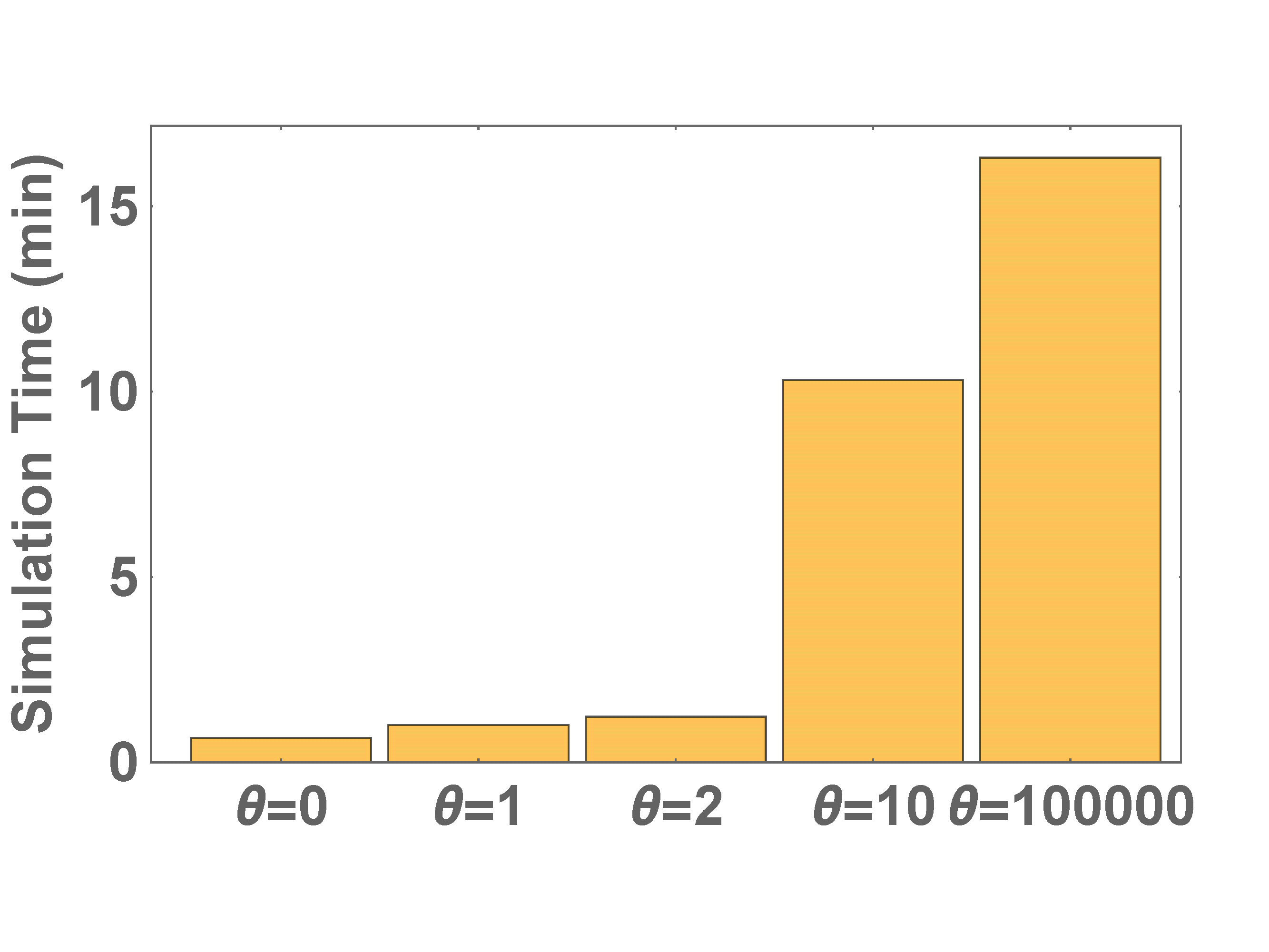}
\caption{Simulation performance as in Figure \ref{fig:MinPerformance}, but with only a quarter of the total number of particles. 
\label{fig:MinPerformance_350}} 
\end{figure}

\end{document}